\documentclass[conference]{IEEEtran}
\usepackage[a4paper,left=3cm,right=2cm,top=2.5cm,bottom=2.5cm]{geometry}
\usepackage{cite}
\usepackage[pdftex]{graphicx}
\usepackage[cmex10]{amsmath}
\usepackage{amssymb}

\usepackage{times}
\usepackage{array}
\usepackage[tight,footnotesize]{subfigure}
\usepackage[font=footnotesize]{subfig}
\usepackage{graphicx}
\usepackage{mathtools}
\usepackage{amsmath}

\def\intersect{{\cap}}

\usepackage{alltt}
\usepackage{shadethm}
\newshadetheorem{comment}{Comment}
\newtheorem{lemma}{Lemma}
\newtheorem{theorem}{Theorem}
\def\R{\mathbb{R}}

\def\boxx{{\vcenter{\vbox{\hrule height.3pt
          \hbox{\vrule width.3pt height6pt
          \kern6pt\vrule width.3pt}\hrule height.3pt}}\;}}

\def\impos{{\;\vcenter{\hbox{\rule{5mm}{0.2mm}}} \vcenter{\hbox{\rule{1.5mm}{1.5mm}}} \;}}

\def\lrarrow{\leftrightarrow \kern-8pt \rightarrow}

\def\Unspec{A_{\rm ?}}

\def\2{\frac{1}{2}}

\def\beq{\begin{eqnarray}}
\def\eeq{\end{eqnarray}}
\def\2{\frac{1}{2}}

\newtheorem{example}{Example}

\newtheorem{principle}{Principle}

\newtheorem{definition}{Definition}

\def\lrarrow{\leftrightarrow \kern-8pt \rightarrow}

\def\frightarrow{\rightarrow \kern-11pt /~~}
\def\reducesto{\simeq \kern -3pt >}
\def\intersection{\cap}

\usepackage{fancyhdr}					
\begin{document}
\newcommand{\strust}[1]{\stackrel{\tau:#1}{\longrightarrow}}
\newcommand{\trust}[1]{\stackrel{#1}{{\rm\bf ~Trusts~}}}
\newcommand{\promise}[1]{\xrightarrow{#1}}
\newcommand{\revpromise}[1]{\xleftarrow{#1} }
\newcommand{\assoc}[1]{{\xrightharpoondown{#1}} }
\newcommand{\rassoc}[1]{{\xleftharpoondown{#1}} }
\newcommand{\imposition}[1]{\stackrel{#1}{\impos}}
\newcommand{\scopepromise}[2]{\xrightarrow[#2]{#1}}
\newcommand{\handshake}[1]{\xleftrightarrow{#1} \kern-8pt \xrightarrow{} }
\newcommand{\cpromise}[1]{\stackrel{#1}{\frightarrow}}
\newcommand{\policy}{\stackrel{P}{\equiv}}
\newcommand{\field}[1]{\mathbf{#1}}
\newcommand{\bundle}[1]{\stackrel{#1}{\Longrightarrow}}

\title{From Observability to Significance\\in Distributed Information Systems}
\author{Mark Burgess\\~\\Aljabr Inc.\\ChiTek-i AS\\~}
\maketitle
\IEEEpeerreviewmaketitle

\renewcommand{\arraystretch}{1.4}

\begin{abstract}
  To understand and explain process behaviour we need to be able to
  see it, and decide its significance, i.e. be able to tell a story
  about its behaviours.  This paper describes a few of the modelling challenges
  that underlie monitoring and observation of processes in IT, by
  human or by software.  The topic of the observability of systems has
  been elevated recently in connection with computer monitoring and
  tracing of processes for debugging and forensics.  It raises the
  issue of well-known principles of measurement, in bounded contexts,
  but these issues have been left implicit in the Computer Science
  literature.  This paper aims to remedy this omission, by laying out
  a simple promise theoretic model, summarizing a long standing trail
  of work on the observation of distributed systems, based on
  elementary distinguishability of observations, and classical
  causality, with history.  Three distinct views of a system are
  sought, across a number of scales, that described how information is
  transmitted (and lost) as it moves around the system, aggregated
  into journals and logs.
\end{abstract}



\section{Introduction} 

The desire to expose the chains of cause and effect that lead to
observable phenomena is irresistible, yet we know that the ability
to observe and interpret systems is both limited and
compromised by availability, relativity, signalling speeds, and even by the noise of
environment.  In computer science, these limitations have largely been
set aside to prioritize other concerns, but in the era of
wide area cloud computing this will become harder to do, as our
ability to observe and assess software behaviours is filtered through
ever more opaque layers of abstraction.

The state of the art in monitoring relies principally on brute force
data collection and graphical presentation. There is surprisingly
little discussion about the semantics of the
process\cite{usenix9315,lisa95125,hellerstein2,stadler3,sekar1,dasgupta1}.
Only recently has there been any serious interest in semantics for 
distributed tracing\cite{opentracing,william3,william4}.
Uncertainties, accrued by sensory instrumentation are left for human
operators to untangle on their own.  With few exceptions, the
literature on logging and monitoring (prior to the present wave of
Machine Learning studies), singles out the design of machinery to
collect data, without due consideration of relevance, accuracy, or
semantics of what is collected (some recent examples include
\cite{seltzer1,chen1,emerald,ifip99371,bigdata,barford_crovella}).
This is regrettable, but not uncommon in technological literatures.
The machinery becomes an end in and of itself, and its knowledge-related
function is subordinated to the prowess of its performance.

Interestingly, the collection of observational measurements from a
distributed system is related to the far-more widely studied problem
of data {\em consensus}, in Computer Science\cite{paxos,raft}. The
latter considers how we may distribute multiple copies of information
over a wide area, with integrity of order--- surely one of the most
frequently revisited problems tackled in distributed systems.  The
fascination with consensus stems from the attempt to cling onto
approximate determinism.  The topic of {\em observability}, on the
other hand, (literally the ability to observe systems, as contrasted with
monitoring which is following what you actually can see) is its
approximate inverse: how can we meaningfully integrate data from
widespread sources into a viewpoint consistent with a single observer,
also with integrity.  It's an issue that has hounded physics and
engineering for centuries.  Like its dispersive counterpart,
observability is a problem dominated by relativistic issues of space
and time. Unlike consistency, it has not been studied with anything like the
same degree of care.

The `measurement problem', as it is known in physics, bedevils every
corner of science in different ways.  We need to ask: is there a
consistent viewpoint that can be arrived at without doing such
violence to the system as to wipe out any other signal.  In this
paper, I want to apply the language of Promise
Theory\cite{promisebook} and Semantic
Spacetime\cite{burgessdsom2005,promisebook,spacetime1,spacetime2,spacetime3,cognitive}
to the understanding and analysis of distributed systems on any scale.

IT tends to favour action over understanding; this has probably led to
the neglect of detailed models for process monitoring. The issues of
how we cope with preferential sampling (which may result in result
bias) and relative scaling of sample populations, for instance, is of
serious importance, but completely absent from monitoring literature.
Statisticians talk about biases and significance of populations, but
the normal state of affairs for any observer, watching in band of the
process (i.e. colloquially in `realtime'), is to see a small random
sample of data whose larger scale significance cannot easily be
assessed without long term studies. Decisions and value judgements
are inevitably based on samples that are small and statistically
inadequate, so this problem cannot be argued away by Central Limit
Theorems and the like.  Observability is a necessary but not
sufficient criterion for understanding. 

None of this is not what monitoring software purports to do---rather
it pretends to offer instant insight, independent of scale.  In this
paper, I'll try to set out some definitions as clearly as possible,
with a view to answering a few basic questions, especially: can we
feasibly collect enough information to enable reversible
reconstruction of process history, thus enabling forensic causal
reconstruction of scenarios past?  I'll show that, if a sufficient
level of information is promised, and agents keep their promises with
sufficient fidelity, then trajectories can be traced reversibly to
causal roots. However, the reconstruction of an agent's ealier {\em state}
from its current state (the tracing analogue of `rollback'), or from
promised data, should be considered impossible in general\cite{jan60}.

I'll focus on the
two necessary conditions for assessing systems, from basic information theory\cite{shannon1,cover1}:
\begin{itemize}
\item The ability to observe remote data with reasonable fidelity.
\item The ability to aggregate and combine remote observations with similar fidelity.
\end{itemize}
These points are never more important than in extended cloud computing,
where data collection systems extend all the way out to the edge of user contact,
e.g. the Internet of Things.

\section{Notation and definitions}

Let's begin with some definitions for the purpose of making more precise
statements.  Promise Theory provides a useful language here, in terms
of promises (or impositions), and assessments.  In a promise theoretic
model, any system is a collection of agents. Agents may be humans or
machines, hardware or software. Usually, agents will be {\em active
  processes}. Agents represent internalized processes that can make
and keep generalized promises to one another\cite{promisebook}.

The generic label for agents in Promise Theory is $A_i$, where Latin
subscripts $i,j,k,\ldots$ numbers distinguishable agents for
convenience (these effectively become coordinates for the agents).  We shall often
use the symbols $S_i$ and $R_j$, instead, for agents to emphasize
their roles as source (initiator) and receiver (reactive).
So the schematic flow of reasoning is:
\begin{enumerate}
\item $S$ offers (+ promises) data.
\item $R$ accepts (- promises) or rejects the data, either in full or in part.
\item $R$ observes and forms an assessment $\alpha_R(.)$ of what it receives.
\end{enumerate}
This third and final stage is the moment at which data can be said to arrive at the receiver.

The details of a physical network are not directly relevant, but the
topology of actual interactions between agents is. It depends on the promises made
between pairs of agents, which therefore serve as documentation of intent. 
An offer promise with body $+b_i$ made by $S_i$ to
$R_j$ is written:
\beq
S_i\promise{+b_i} R_j,
\eeq
where the $+$ refers to an offer of some information or behaviour (e.g. a service).
This is a part of $S_i$'s autonomous behaviour, and the promise constrains only $S_i$.
$R_j$ may or may not accept this by making a dual promise, marked $-b$ to denote the
orientation of intent:
\beq
R_j\promise{-b_j} S_i.
\eeq
If both of these promises are given, and kept, then influence in the form of 
vital information about the body $b$
will pass from $S_i$ to $R_j$. In general, the offer and acceptance may not match
precisely, in which case the propagated information will be the overlap (mutual information)
\beq
b_\intersect = b_i \intersection b_j,
\eeq
in the manner of mutual information\cite{shannon1,cover1}.
I'll suppose that modern systems are cloud computing systems.
The elementary agents of cloud computing are {\em processes}, any of which
may express promises about state and services. Processes are hosted at agent locations 
$A_i,S_i,R_i$ etc.

I use the following nomenclature for message agents $M$:
\begin{itemize}
\item $E_\gamma$ is an event, for example $L_\gamma \subset E_\gamma$ may be a 
line of information reported in a log or journal. Greek indices $\gamma$ label information agents
successive packets, i.e. $\gamma = 1, 2, 3, \ldots$.
\item $\{E_\gamma\}$ or$\{L_\gamma\}$ refers to a collection of such events or lines.
\item $S_i,R_i \in A_i$ refer to processes running on computers.
\item $\{S_i\}$ refers to a collection of sources, etc.
\item $A_i$ refers to a process checkpoint in some kind of dataflow,
which has its own interior event log and counters. Checkpoints 
typically make promises about their identity, location, local counter values, and intent to pass
on data in the form of packets $P_i$, with some promised order.
\item $P_i$ refers to a data packet passed between checkpoints agents.
Packets typically make promises about their identity, data content, schema, and
type.
\end{itemize}
Latin indices therefore label locations, and Greek indices label events at the same location.

\section{Defining the problem}

\subsection{What is intended and what is promised?}

There are two ways in which we use data to interrogate a number of processes:
\begin{itemize}
\item {\bf Tracing}: (`During') ---in band observation, in which data are sampled {\em intentionally}
  and recorded as a process unfolds to maintain `situation
  awareness'. e.g. the ECG or life monitor approach to medical monitoring.

\item {\bf Diagnosis}: (`After') ---out of band forensic 
reconstruction of a system using data one can find after
an incident, where intent to comprehend kicks in only after the event has occurred:
e.g. the post mortem approach to medical investigation.
\end{itemize}
Most users will try to combine these approaches, paying attention
mainly when significant events occur. The automation of alarms (usually
based on simple-minded absolute thresholds) tells human operators
when to pay attention, at which point they have to rely on what
has been traced. The promise to maintain awareness is an expensive one,
and we rely heavily on our skills of reconstruction after the fact.

\subsection{Three perspectives about scale and relativity}\label{storytypes}

\begin{figure}[ht]
\begin{center}
\includegraphics[width=6cm]{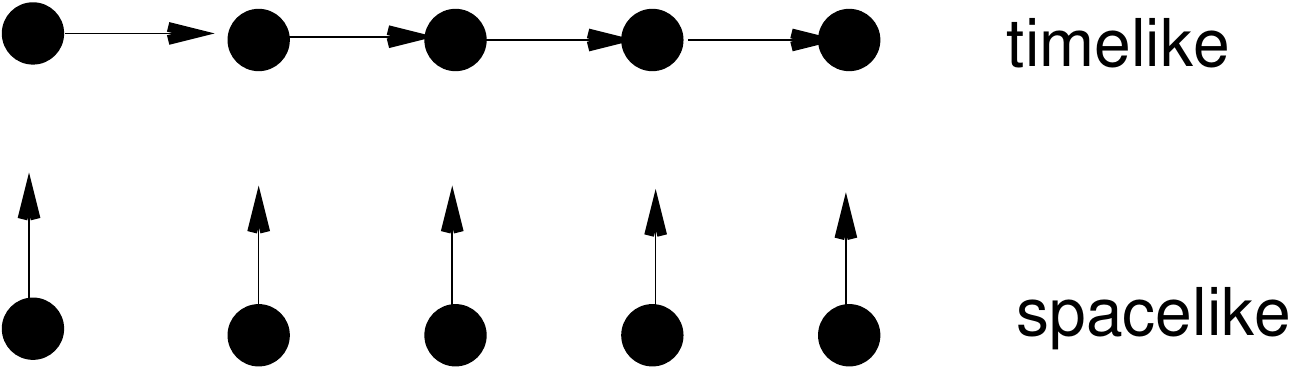}
\caption{\small Space and time as agent parallelism and serialism respectively.\label{st1}}
\end{center}
\end{figure}

Distributed processes are composed of agents that pass information in
space and time (see figure \ref{st1}).  Messages or events propagate
from one agent to another, and we consider the arrival of such
information to be an advance in the `state' of the distributed system,
which is what we mean by the {\em proper time} of the process. Events
that occur in parallel, as separate logical locations, know nothing
about one another---they are causally disconnected and lead
independent lives. The time on the wall clock or system clock is not a
`proper' time, as we'll see below\footnote{In Einsteinian relativity,
  the term proper time is reserved for the time experienced by an
  observer about its own states, so I keep to this convention here.}.

\begin{figure}[ht]
\begin{center}
\includegraphics[width=6cm]{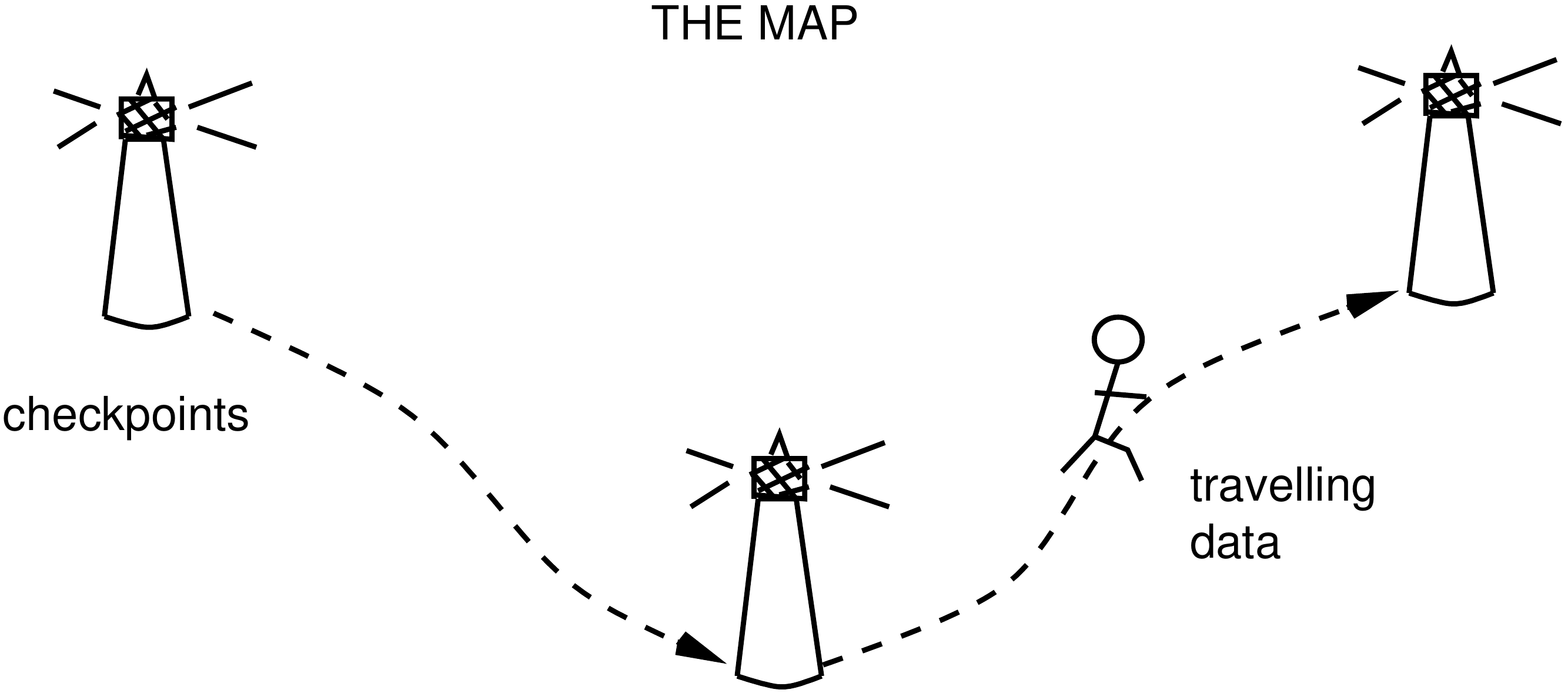}
\caption{\small 3 Views. Travelling passport documents, versus logs of entry and exit from a checkpoint, versus
the map of checkpoints and routes.\label{views}}
\end{center}
\end{figure}

There are three kinds of story or explanation we want to be able to tell about
distributed systems (figure \ref{views}):
\begin{enumerate}
\item The data traveller log.  What a travelling data packet experiences
  along its journey, e.g.  which software including version handled it
  and in what order?

\item The checkpoint visitor log, from key signposts around the data processing landscape.
The log of what each checkpoint along the journey saw, i.e. which data packets
passed through the checkpoint and what happened to them?

\item The map of combinatoric intent, i.e. the relationships between
  invariant elements and concepts, including the topology of
  checkpoints and influences, the types of data passed between them,
  significant occurrences, and so forth; i.e.  the semantics of the
  data, software, and invariant qualities and quantities that
  summarize the processes within the system's horizon.
\end{enumerate}
These viewpoints require separate data collections.
Present day logging systems focus almost entirely on the second of these.

In Promise Theory, one reduces a system to a collection of agents,
their promises, and their assessments.  Agents include the checkpoints
from which data emerge and are collected.  A second layer of agents
comprises the data packets that are transmitted. The promises made by
these agents include communication, data compression, speed, and
integrity.  They may include data formats and ordered protocols.  We
equip different agents in the system with promises to report the
information available to them to observers. I shall not be concerned
with matters of authorization and permission in this paper, but rather
focus on the difficulties experienced by those who are promised
information.

\subsection{Diagnosis}

It's up to an observer to infer something about the state and history
of a system, based on what is observed. This is not as straightforward as
software systems have come to assume, especially as cloud computing
pushes the limits of observability. At some point, this reconstruction
involves a form of reasoning---not necessarily rigid logical reasoning,
but at least a process of joining dots into an acceptable story.

\subsection{Diagnostic messages}

Process tracing is a simpler problem than reasoning, because it can be
constructed as a purely Markov process---at least in principle.
Tracing is the construction of a totally ordered path through a set of
agents.  Reasoning, on the other hand, involves semantic relationships
between clusters of agents that may be considered to have an invariant
meaning, and it may combine several traces into a satisfactory
explanation.  According to the definitions in
\cite{spacetime3,cognitive}, I'll simply define the following:

\begin{definition}[Reasoning]
Reasoning is a search over a graph of ordered conceptual relationships.
\end{definition}
This pragmatic and unconventional definition might offend some logicians, but it's 
closer to what humans call reasoning than a definition based on mathematical logic.  A
few common issues crop up in diagnostics:
\begin{itemize}
\item The predictability of agents' behaviours.
\item The distinguishability of agents and data messages.
\item Loss of information due to mixing of origin sources.
\item Reordering of information due to latency.
\end{itemize}

The problem with the first two kinds of story in the list, is the lack
of a deterministic and universally defined order between the
transactions of `event driven' processes at separate source locations.
The extent to which we can write down spatially invariant orders,
process summaries, etc, which may be expected to persist over a
timescale useful for prediction, is the essence of the difficulty in
tracing causal history\footnote{The main approach to determining
  spacetime invariance in science is by the use of statistics
  (aggregation or `learning'): by accumulating multiple samples, we
  hope to separate what is quickly varying (fluctuation) from what is
  slowly varying (trend).  Persistent concepts are what remains during
  or after a process of learning has separated these processes.}.

\section{Observability of messages}

From the foregoing, we can define the concept of observability between pairs of agents.
It does not make invariant sense to speak of `observability' without some qualifications,
so we can be more precise:
\begin{theorem}[Observability of $X$ at $S$ by $R$]
A range of promised set values $X$, sourced from an agent $S$ is observable by an agent $R$ if
and only if:
\beq
\pi_+ : S_i &\promise{+X_i}& R_j,\\
\pi_- : R_j &\promise{-X_j}& S_i.\\
X_j &\subseteq& X_i
\eeq
\end{theorem}
Note that the criterion for observability is not a deterministic
guarantee the ability to obtain a value on demand. It is essentially
a property of an information channel, in the Shannon sense. There is only a
finite probability of all these promises being kept, which makes
observation a fundamentally non-deterministic process. There are many
impediments to keeping promises in practice, not least of which the
the law of intermediaries\footnote{See reference \cite{promisebook}.
  The law of intermediaries basically says that intermediate agents
  cannot be relied upon to faithfully transmit promises or intent, because all agents
  are fundamentally autonomous.}.

By assumption, agent's are autonomous and each plays a role in
the collaboration required to exchange the
information involved in monitoring.  The definition of observability
illuminates a basic dilemma in monitoring: the autonomy of agents in
any distributed process (i.e. their causal independence) means that
there is fundamental uncertainty about the process of observation not
just its outcome. Causal independence is the very definition of a
random variable.  A source of signals may believe it does all it can
to ensure correct transfer of information, and that any problems lie
in the delinquencies of the receiver; meanwhile, the receiver believes
it does all it can and trusts the source and the network in between
implicitly to report with complete fidelity. The assessment of $X$ by
$R_j$ (denoted $\alpha_j(\pi_+)$) is still a function of $R_j$'s
access and capabilities at any given sample, and may be subject to
environmental interference.

\subsection{Preliminaries about intent}

Most technologists believe that, if they design without `bugs', they can achieve
whatever outcome they desire, given sufficient resources. This is not a scalable view,
so we need to be more cautious.
In the standard model of queueing theory \cite{kleinrock1}, data 
are produced by a source $S_i$, at a rate $\lambda_i$ messages per unit time,
and may be processed by a receiver $R_j$ (sometimes called a server) at a rate
$\mu_j$. The queue is unstable and grows out of control
as the traffic density $\lambda_i/\mu_j\rightarrow 1$.

The terminology pulling and pushing data are often used about
attitudes to causal {\em intent} in communications. These lead to some confusion,
so it's worth mentioning them. I define these in
accordance with the same information channel principles.
\begin{definition}[Push]
  A method of communication in which a source agent $S$ imposes its
  messages onto a recipient $R$ without invitation.  \beq S
  \imposition{+M} R.  \eeq In data signalling, packets may be carried
  over a wire, enter a network interface and be queued up for sampling
  by a receiver, before the receiving process is ready to accept them.  
This is imposition. The receiver may then promise to sample (-) the
  messages from the shared queue. The channel flow is thus controlled
  by the sender.
\end{definition}
\begin{definition}[Pull]
A method of communication (sometimes called publish-subscribe) in which
a receiver invites a source to provide a certain quota of messages
for sampling via an agreed channel. 
\beq
R &\promise{-M}& S
\eeq
which is then promised by the source
\beq
S &\promise{+M}& R
\eeq
The flow along the channel is thus controlled by the receiver.
\end{definition}
Pull is always the fundamental `last mile' stage of a sampling
operation; it may involve active polling of the queue to match
timescales that satisfy the rigours of Nyquist's theorem. It optimizes
message transfer according to the downstream capabilities.  Push
driven systems are sometimes associated with {\em reactive} or {\em
  event driven} systems---though this can be misleading. Push and pull
are effective on different timescales. Push (notification) is useful
in connection with small signals when source data are sparse and a
receiver needs a short wake up message to collect a package from the
source, enabling it to conserve resources.  Push therefore provides
non-redundant information when arrivals are sparse or infrequent, on
the timescale of the receiver's sampling\cite{remi3}. Pull systems
make more efficient use of queue processing, by utilizing the information
about autonomous capacity to balance load.

\subsection{Events and sampling}

The concept of events plays a major role in the language used
for monitoring and data flow in IT, but is seldom defined. Let's define it here 
in a way that respects information theoretic transfer. Information is
only `arrives' somewhere when it is sampled.
\begin{definition}[Event]
A discrete unit of process in which an atomic change is observed or sampled.
\end{definition}
We often imagine processes being driven by a flow of events, like a
stream\footnote{This illusion of a flow is maintained by a gradient of intermediate agents
that accumulate samples in buffers.}. Again, in terms of sampling, this amounts to
the following:
\begin{definition}[Event or message driven agent]
An Event Driven Agent $R$ makes a promise conditionally on the sampling
of message events $M$ from a $S$, with an average rate $\lambda$:
\beq
R \imposition{+E | M} O\label{emp}
\eeq
i.e. $R$ can promise an observer $O$ that it acknowledges an event $E$ on receipt
of a message $M$. By Promise Theory axioms, this assumes the prior promises:
\beq
S \imposition{+M | \lambda} R &\text{~~\rm or~~}& S \promise{+M | \lambda} R\\
R &\promise{-M | \mu}& S
\eeq
\end{definition}
Notice that by using the term sampling here, we do not take a position
on whether messages were imposed by pushing from $S$ to $R$, or
whether $R$ reached out to $S$ to pull the data. These distinctions
are irrelevant to the causal link that results from the message
policy. Data are not received until they are sampled by the receiver.  Note that there
is no timescale implied by the conditional promise in
(\ref{emp})---the definition of `immediate' or `delayed' response is
an assessment to be made by the observer $O$.

The concept of reactive systems has been usurped by a specific
industry initiative \cite{reactive}, so it makes sense to follow in
the spirit of that:
\begin{definition}[Reactive agent]
An event driven agent that promises to keep its behaviour within certain 
constraints relative to the sampling of events.
\end{definition}
A simple example of these principles is to explain 
data flow systems like Data Pipelines. These are scaled
combinations of reactive components.
\begin{example}{Pipelines and Petri Nets:}
  Data processing pipelines are hybrid networks of Event Driven
  Agents, Reactive Agents, and Service Agents, that promise to behave
  as an Event Driven Agent collectively as well as component by
  component.
\end{example}
A simple fact of queueing theory, embodied as a principle of autonomy
in Promise Theory is to note that: no amount of pressure or coercion
will make an agent process data faster than its maximum rate $\mu$.

\subsection{Missed and dropped samples}

Observations inevitably get lost in any scientific enterprise. In
empirical science this contributes to `error bars' or uncertainties in
counting of measurements---but not usually to semantics of interpretation.
Interpretations are expected to be stable to such small perturbations.

Reasoning in IT has its historical origins from mathematical logic and
precision: the avoidance of doubt. But doubt is a central part of
tolerance in systems. If we observe and inspect systems, we need to do
so in the framework of a stable intent that overrides random
fluctuations in measurement\footnote{We see the effect of `populism'
  in society today, when intentions follow unstable polls instead of
  convictions.}.  Monitoring and measurement serve no actionable
purpose unless there is already a policy for behaviour in place.
Ashby's model of requisite complexity or `good regulator' in
cybernetics\cite{ashby1,ashby2} summarizes how matching information
with information on the same level is required when there is no
intrinsic stability in a model by which to compress such fluctuations.

\section{The Time Series Model}

Our received view of time---as a river of events that moves everything
from past to future at the same rate---is a side effect of living
in a rather slow world, which is close to us, and which we see with no
perceptible delay. In IT, we cannot rely on this privileged view,
and we need to rethink time by going back to the basics of how we
measure it.

\subsection{Events, clocks, and proper time}

In an information theoretic sense, an {\em event} is an observation of
change in data sampled from a source\cite{shannon1,cover1}.  In the
Einsteinian sense, this signal is a tick of a clock that an observer
samples. When the tick originates from within a process (e.g. a CPU
kernel tick), this defines a notion of `proper time' for the local
process, indicating an advance in the state of the process.  When
there are multiple agents involved, working together, the language one
often speaks of `vector clocks' in IT, referring to Lamport\cite{vectorclocks}.

Other agents, external to a ticking process, may
observe changes in it differently, either because they lack access to
observe the changes or because the sampling of the changes require
intermediary processes like message passing to propagate the changes from
source to receiver.  Thus the proper time experienced by an agent may
not correspond to the exterior time generated by the sampling of
remote events.

\begin{example}[Thunder and lightning]
  When lightning strikes, observers in different contexts see and hear
  it at different times. Observers very close, that cannot sample
  faster than a certain rate, may not be able to discriminate a
  difference between the flash and the thundercrack.  Light travels so
  fast the few agents can detect a delay in the signal, so they
  conclude that the flash occurs as `the same time' (during the same
  sample).  But sounds travels more slowly, so agents at different
  distances can discriminate the time at which the sound reaches them.
  If they synchronize their watches using light, they will measure
  different times for the sound---but the event happened due to a
  process that took place in a single location, over a tiny fraction
  of a second. What observers sample is not always a high fidelity
  representation of what happened at the source.
\end{example}

Using the language of Promise Theory\cite{promisebook}, we can define
time from two perspectives.
For convenience, we'll make an identification between the concept of an event $E_\gamma$ and
a line transaction $L_\gamma$ in a system log, or a data point recorded in a timeseries
database $D_\gamma$:
\begin{lemma}[Events count time]
The emission of an event or `log line' $E_\gamma=\{L_\gamma,D_\gamma\}$ is a tick of interior time clock.
\end{lemma}
This should be obvious, as events are changes that get noticed.
We can now define interior (proper) time and exterior (relative) time:
\begin{definition}[Interior time of process $S_i$]
An independent count of ticks originating from within a process $S$,
cannot be observed by any exterior agent $\Unspec$, unless promised and reported:
\beq
S_i \promise{+\text{tick}_i} \Unspec.
\eeq
\end{definition}
Interior time is the image of processes that originate within the boundary of agent $S_i$.
At scale, we can consider superagents of any scale, so interior time scales and changes
in meaning according to our definition of local.

\begin{definition}[Exterior time of process $S_i$]
  An independent count, by a remote receiver $R$, of promised ticks
  (observed and aggregated from any number of sources on a watchlist) that increases
  for each sampled event arriving from a exterior process source $S_i$.  \beq
  S_i &\promise{+\text{tick}_i}& R\\
  R &\promise{-\text{tick}_i}& S_i\\
  S_i &\promise{+\text{tick}|\text{tick}_i}& \Unspec.  \eeq
\end{definition}
Exterior time is attached to remote processes that may originate on
any scale.  The recipient $R$ that samples events may itself be of any
scale, with associated loss of certainty about the definition of its
interior clock counters, but ideally $R$ would use a single source
from an elementary agent, for precision. 

On the timescales of computers, in our daily lives, this sounds
straightforward, but the processes that calibrate our normal idea of
time (the system clocks) are not faster than the sampling processes we
are trying to discriminate by.  This leads to a breakdown in the
normal assumptions of universal time for all, and forces precision
agents to go back to basic definitions of time in order to trace
processes in band.

\subsection{Clocks at different scales}

We cannot avoid the effects that scaling has on clocks. Even atomic clocks may
not be considered atomic, in the transactional meaning, on the scale
of subatomic processes.  The lesson that Einstein taught us is that
processes need to embody their own clocks, as single reference sources
of truth.

\begin{example}[System clock, e.g. Unix]
The system clock, provided by most operating systems
derives from a shallow hierarchy of exterior time services,
based on processes that promise {\em approximate} alignment.
The clock timer $C$ is an independent agent, which
promises a counter (UTC) to processes $P_i$,
\beq
C &\promise{+\text{UTC}}& P_i\\
P_i &\promise{-\text{UTC}}& C
\eeq
Using this as a conditional dependency, processes can then promise
timestamps based on the interior counter
\beq
P_i &\promise{+\text{timestamp}|\text{UTC}}& R.
\eeq
Note that the coordination between duplicate redundant clocks is weak. A time
service like NTP, provided by agent $N$, may be used to periodically align independent clocks
at a layer of the hierarchy above each system clock:
\beq
N &\promise{+\text{UTC}_\text{NTP}}& C\\
C &\promise{-\text{UTC}_\text{NTP}}& N\\
C &\promise{+\text{UTC}|\text{UTC}_\text{NTP}}& P_i\\
P_i &\promise{-\text{UTC}}& C.
\eeq
Each clock is independent, so it is only meaningful to compare
two timestamps from the same clock. Moreover, the relationship
between timestamps and process ticks is indeterminate, since
process ticks are halted relative to the system clock during
timesharing. The use of timestamps in network protocols should
be considered unreliable, and only for round trip comparisons.
\end{example}

\begin{figure}[ht]
\begin{center}
\includegraphics[width=7cm]{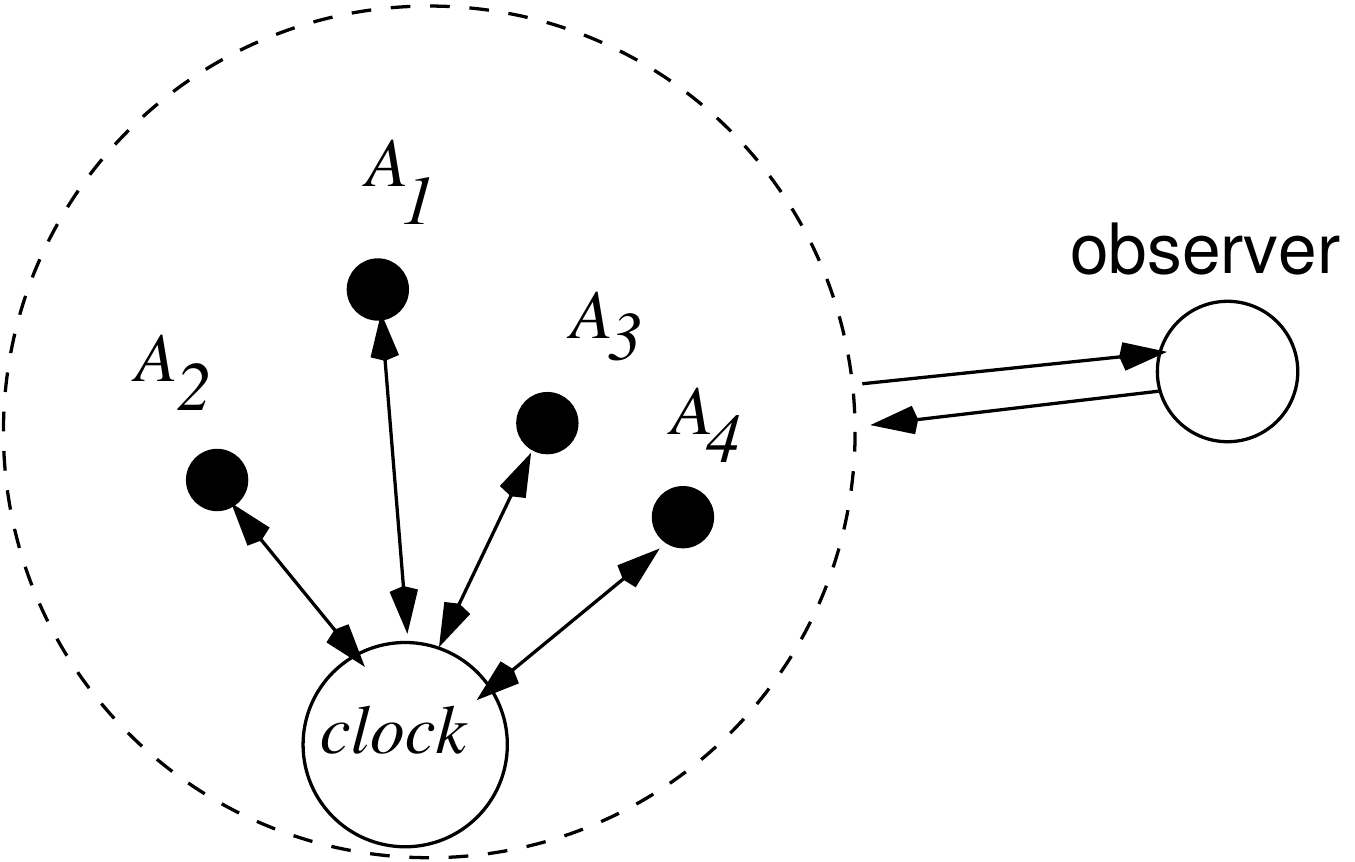}
\caption{\small A single agent, with a reference clock
can be scaled into a superagent provided the agents
within promise to coordinate their behaviours. Thus
a collective formed from independent sources can act
as a single reliably ordered source, but at a cost
growing like $N^2$ in the number of agents. \label{superagentclock}}
\end{center}
\end{figure}

\begin{example}[Monotonic counters]
  Interior process time can be obtained by incrementing a counter by
  an atomic operation. A process $P$ passes a value $v$ to a counter $C$, which is a
  persistent variable, and promises to increment it as certain
  milestones are passed: 
\beq 
P &\promise{+v}& C\\
C &\promise{-v}& P\\
C &\promise{+t=(v+1)|v}& P\\ 
P &\promise{-t}& C.
\eeq
\end{example}

If independent agents need to coordinate their clocks, they can
build on a single source of truth, by appointment to the role (see figure \ref{superagentclock}),
essentially transforming interior time into exterior time.
\begin{lemma}[Interior consensus of clocks]
  The promise to share interior time $T_i$ from $S_i$, to an agent
  $R_j$, with interior clocks $T_i$ is equivalent to the problem of
  data consensus between clock ticks.
\end{lemma}
This suggests that clock synchronization by voting in band
(`realtime') will lead to a significant delay in the rate of 
time that can be promised as agreed ticks by an entire superagent.
This increased `mass' of agent clocks will slow the rate observable
by an outside sampler.
\begin{lemma}[Aggregation of clocks]
The aggregation of multiple sources of interior time $T_i$ from $S_i$, by an agent $R_j$,
with or without consensus, is not one to one with the interior time of the receiver.
\end{lemma}
The proof of this may be seen from the Law of 
Intermediate Agents\cite{promisebook}, which tells us that, if there are agents in
between the source and receiver (which is nearly always the case),
then no promises are transferred automatically. We need a chain of
delivery promises to form an expectation of what we are seeing.  The
outcome of this is that every agent may see a different arrival order,
assuming that it can distinguish between data transmissions.
This is a well-known result.
Conversely:
\begin{lemma}[Promised Order Propagation]
The order of a sequence of data, from a single agent, can be promised
by virtue of a single clock or counter.
\end{lemma}
Note that this does not imply that order will be preserved, only that
there is a set of promises between sender and receiver that can be
made to transfer the order information (e.g. by numbering packets).
This is well known in `reliable' data communications, like TCP.
Without proof, let's acknowledge that the relative order of
data can indeed be transferred reliably between a sender and a receiver, if
there is a promised order at the source (figure
\ref{superagentclock}). This requires the introduction of a
co-dependence between sender and receiver, and a detailed explanation
has been given in \cite{paulentangle}. Examples include the well-known
TCP protocol, and other more exotic variants.

This does not imply that data are necessarily observed in the same order
by source and receiver. Once data leave the agents that are entangled
in this way, the promise of order is not preserved, because
all agents are causally independent. 
\begin{lemma}[Promised Order Propagation]
Data exchanged without conditional sequence promises may
not be sampled in the same order as they were promised.
\end{lemma}
The implication of this lemma is that predictable coordination of
sequences as invariant features between agents is expensive and
unreliable: it does not happen unintentionally, without chains of
interdependent promises.  
This is a rather damning result for monitoring that relies on timestamps
for its depiction.

Order promises can be kept by labelled
(sequence numbering) or by waiting in lock-step for changes one at a
time.  The independence of agents, and our inability to make a promise on
their behalf, means that data passing through multiple
intermediaries are {\em independent deliveries}. Even a single agent
cannot be forced to deliver data in order, unless it has promised (fully intends) to
do so in advance, with full observability of the payload (and a receiver
that can sample at the Nyquist rate\cite{cover1,spacetime3}). Expecting
clusters of agents to preserve order, over possibly parallel routes is
even more unlikely, without prior intent. This can be expressed by
saying that unless there is a single clock that determines when
packets will be sampled by a receiver, the order will not be
preserved. The default is that an incoming queue of samples serializes
them in a random order, without a surviving chain of dependency.

We can arrange for such a single clock to be authoritative (like a
shared memory counter), but this requires agents to make promises to
abide by the order, which in turn requires cooperation from end to end
throughout a channel, to preserve identity serially and atomically
(one agent sampled exclusively at a time).
Another approach to agreeing about time is to bind clocks 
in lock-step to form a
co-dependent relationship between agents, known as {\em
  entanglement}\cite{paulentangle}. This is used, for example, in
TCP's SYN-ACK protocol. It promises
synchronization at a possibly unbounded cost in terms of interior time ticks. 

\begin{lemma}[Promised Order Propagation]\label{agglem}
  The intended order for events originating from more than one $S_i$
  may only be promised by interior cooperation at the source, and
  assessed uniquely by an agent $R$ with observational capacity
  according to the Nyquist law.  Each rescaling of aggregated time
  ordering introduces new uncertainty according to an observer's
  clock.
\end{lemma}
There is no unique intent, for a collection of autonomous agents,
unless the multiple sources subordinate themselves by cooperation to a
single agreed order, but any attempt to coordinate between the agents
(and thus act as single superagent making a common promise) would
result in a change in the ticks observed by $R$, unless the sampling
resolution of $R$ is much less than the exterior time needed to assess
interior latency of interactions for agreement.

The conclusion of these extended remarks is that there is no single
clock by which to define the order of events between different hosts.
This is esssentially because unrelated processes have no common time.
The whole idea is meaningless. What observers often seek is a picture
according to their own sense of time (observer time) that integrates
different processes into a picture of the moment as they perceive it.
Alas, that impression cannot easily be reconstructed later, even perhaps
with a detailed `post mortem', as it relies on anchoring to out of band
processes that were not measured.
If we introduce a single source of time for a collection of hosts, by
forming a superagent (with all necessary interior cooperation), each
agent within, we can define a single reference time, but it is not the
proper time of any process.

\subsection{Rule of thumb about time}

\begin{example}[Common assumptions of system time]
A commonly held belief is that, in interactions like network
protocols, we might define time in a number of way.
\begin{itemize}
\item {\em The `actual' time}: there is a single source of truth, by
  international convention, which is the official value of UTC. This
  time standard exists, but is only obtainable with latencies that
  render it approximate.  Through a hierarchy of services, like NTP,
  local system clocks promise to approximate this time and to count
  independently on their own at approximately the same rate. These
  rates cannot be verified, so in practice the closest we can obtain
  is the current value of the local system clock, which belongs to
  localhost.

\item {\em The observed time}: This is a timestamp rendered by
  sampling the system clock, so it is relative to localhost's assumed
  time standard and has no significance beyond the agent that sampled
  it. The observed time may not be monotonic, for example if clock
  drift corrections occur in between samples of the clock. System time
  may therefore go backwards or forwards at random, over extended
  processes.

\item {\em The publication time}: Timestamps may be shared between processes,
or recorded, incurring additional processing delays. The resolution
of a timestamp may be quite low, allowing processes to absorb processing
delays, but publication times are always later than the timestamps
they promise, e.g. the timestamp when a log entry is written is always later
than the timestamp of the log entry.

\item {\em The receipt or sampling time}: If timestamps are shared
  between agents, e.g. in recording data in a log, or transmitting
  data across a network, the published timestamp belongs to the sender
  $S$'s clock, and the receipt time belongs to the receiver $R$'s
  clock. These two times are causally independent and their comparison
  is strictly meaningless. If all clocks promise approximate
  alignment, the difference between published and received timestamps
  may promise an accuracy whose uncertainty is approximated by the
  Pythagorean average of the uncertainties of the two timestamps at
  $S$ and $R$.
\end{itemize}
Clearly, no protocol (except NTP) passes information about its clock time
uncertainties, so network time falls foul of the Intermediate Agent law.
\end{example}

\section{The role of timescales for predictability}

The purpose of monitoring is to be able to explain behaviour and even
predict problems in advance.  Without predictability, monitoring is
little more than somewhat arcane
entertainment\cite{usenix9315,lisa95125,hellerstein2,stadler3,sekar1,dasgupta1}.
One assumes that, by learning about the past or by building a
relationship with system behaviours in band, we are able to predict
something about the future behaviour. This, in turn, assumes a stability under the
repetition of patterns.
\begin{definition}[Predictability]
A system that has stable and repeated observable behaviour, on a timescale much
greater than the sampling rate, may be called predictable.
\end{definition}
It's, of course, paradoxical that the time when most users want to
monitor systems is when they are {\em least predictable} and providing
observations of no value.

\subsection{Separation of timescales}

We can make another observation about what happens in
interactions.  The principle of separation of timescales is a design
principle for interacting systems, based on the observation that
dynamical influence causes timescales for change to mix. In earlier
work, I've referred to this as the most significant principle for
engineering---more important than the separation of concerns based on
semantic (functional) separation, such as data normalization or
`class', which is the norm in Computer Science.  Briefly, it says:
\begin{principle}[Separation of timescales]
  Functional systems modularize robustly and effectively when
  processes with different characteristic timescales are weakly
  coupled.
\end{principle}
By `robust', we refer to `stability'\cite{treatise2}.
This principle makes a connection to the related problem of data consensus,
which is a strong coupling regime that maintains data consistency over
average timescales.

\subsection{Dynamical coupling defined}

The foregoing assertions can be justified by looking at what
coupling strength means for interacting agents\footnote{The increasingly
pervasive language of Complex Adaptive Systems leads to assertions about
strong and weak coupling, but we need to be able to define those things
to use them.}.
Phenomena that promise changes on very different timescales interact
only weakly and can therefore be treated as logically separate. By
contrast, agents that promise couplings on the same timescale may
influence one another and therefore belong to the same class of
phenomena.  In terms of the foregoing definitions, we can state the
meaning of separation more strongly, as a theorem:

\begin{theorem}[Separation of causal influence]
As the ratio of timescales becomes large $T_R \gg T_S$,
the effective coupling tends to zero
\beq
e \rightarrow 0.
\eeq
tends to zero (weak coupling).
\end{theorem}
To prove this, suppose a series of partially ordered events at an agent $S$ yields a
series $E_\gamma$, $\gamma = 1,2,\ldots$.  Suppose a source agent $S$ transmits the events,
which are aggregated into superagents $E^{(n)}(E_\gamma)$ of dimension
$n$, by the receiver agent $R$, \beq
S &\promise{+ E_1,E_2,\ldots E_n}& R\\
R &\promise{- E_1,E_2,\ldots E_n}& S\\
R &\promise{+\alpha_E | E_1,E_2,\ldots E_n}& \Unspec\label{qq} \eeq so
that the dimension of the information is reduced by a factor of $n$ by
$R$: \beq
\alpha(E_1,E_2,\ldots E_n) &\rightarrow& \R\\
\Big| E_1,E_2,\ldots E_n \Big| &=& n\\
\Big| \alpha_E\left(E_1,E_2,\ldots E_n\right) \Big| &=& 1 \eeq The average
time between events, as assessed by $R$'s clock, may be denoted \beq
T_S &\simeq& 1\\
T_R &\simeq& n.  \eeq So $R$ assesses $S$'s timescale to be 1 and its
own timescale to be $n$: \beq T_R \ge T_S.  \eeq Thus the average
interarrival times for the queue in (\ref{qq}) $\lambda_R \sim
1/T_R$, etc, satisfy: \beq \lambda_R \le \lambda_S \eeq and the
effective influence, in fraction of messages received compared to
messages sent is expressed by a coupling constant: \beq e \sim
\frac{\lambda_R}{\lambda_S} \le 1.  \eeq

In a strongly coupled system $e \rightarrow 1$, timescales
converge to the shortest timescale of the interacting parts, making
systems busier and more work intensive.  The utility of this
observation is that, if one separates causally independent parts of a
system into superagents that make weaker promises to one another, any
observed correlation between phenomena, that exceeds expectation, can
be considered coincidental or potentially faulty. This can be detected
by a change in the proper time event rate, measured by some agent
within a system.  This principle therefore has significance to the use
of observation for detecting faults and design flaws in systems. It
tends to maximize the signal to noise ratio between promised and
non-promised behaviour\cite{treatise2}.

\subsection{Coupling strength, memory, and consensus}

The concept of knowledge is already more uncertain in a distributed
system than in a local system with random processes.  Lamport's papers
about seeking the homogeneity of data sources is effectively a monitoring
problem in reverse. A collection of agents monitors one or more sources
and tries to equilibrate the knowledge they promise. Data consensus
is a conditional promise of policy-determined values (called quora), based on inputs reported
from sources $S_i$:
\beq
S_i &\promise{+E_\gamma}& R\\
R &\promise{-E_\gamma}& S_i\\
R &\promise{+\text{Quorum}(E_\gamma)\;|\; E_\gamma}& \Unspec.
\eeq
This strong coupling, represented by strong dependency on 
data from a complete network of dependencies, 
demonstrated that time and order of events are fundamental obstacles
in a system of distributed computers in which observation has a finite
latency (usually agents that are spatially
separated)\cite{vectorclocks,paxos,raft}.  The topic touches on the
relativity of simultaneity, and how to make sense of differing views
about what causes what.

The relationship with time is revealed by the `FLP result'\cite{flp},
which exposed the essential impossibility of consistent distributed
knowledge in an uncertain `asynchronous' environment.  
In an asynchronous message-passing system, source or delivery agents
may delay messages indefinitely, duplicate them, or deliver them
out of order. In other words, there is no fixed upper bound on how
long a message will take to be received. A consensus policy promises:
\begin{itemize}
\item All trusted nodes promise the same result (a non-local agreement).
\item All trusted nodes will eventually promise a result.
\end{itemize}
Some approaches to working around the 
limitations of asynchronicity play with strong synchrony promises in order to eliminate
these uncertainties\cite{paulentangle}.

In an asynchronous interaction, each agent's proper time may be used to
define `timeouts' to receiving data to keep a process from waiting for
ever for strong dependencies. Timeouts are a workaround that weakens
the effective coupling strength of an interaction, by effectively measuring
latency in interior time.  There is no unambiguous meaning to a timeout, except the
presence of a potential fault. Latency (round trip time) is the only
covariant measure in a relativistic system, because it's one of the
few measures that has a purely local meaning.

In Promise Theory, the intermediate agent theorem is the analogue of
that result: it says that whenever you rely on agents that are not
yourself, to acquire or deliver information, it no longer promises
what its originator intended.  And if a remote agent promises
something, but doesn't promise it to you too, all bets are off and
there is a quadratically growing cost of verifying.
In monitoring, we are not usually interested in a majority view, rather we
are interested in what happened specifically at what we believe was the
certain place and time of origin (though this is also subject to uncertainty).
We are sometimes interested in a statistical view (which is not the
same as a consensus view, because it admits and even measures the
statistical uncertainty of variations. If certain nodes lie (sometimes
called Byzantine behaviour) we want to know about it, not merely cover it
up.  

There is clearly overlap in the concepts of distributed information,
but monitoring seeks a picture of actuality, rather than a cover-up
operation to brush uncertainty under the rug of consensus.  Software
Engineering therefore has a conflict of intent with monitoring: it
wants to assure complete dependability (promises always kept) by
invoking protocols `in band', whereas monitoring is trying to expose
when promises are not kept, to `out of band' human observers.  These
are the issues we need to deal with in describing observability.

Readers may feel that the problem of distributed ordering has been
solved by distributed consensus systems like Paxos and
followers\cite{paxos,raft}, but this is not the case.  Consensus
systems do not promise the source order of observations, but rather
an average order by which observations are reported, which is a
policy decision.

\subsection{Memory processes}

It should be clear that memory is required to stabilize values from
multiple sources. To integrate results from several sources, and to
replace then with an agreed result requires temporary memory, over the
necessary clock ticks of proper time---at least as much memory as
there are source dependencies for each outcome.  The role of memory
and its reliability also play a role, but I don't want to discuss that
here. In most IT systems, memory unreliability is negligible.

\section{The Time-Series model}

Given a stream of trusted values reported by agent interactions, the
usual response is to try to build a timeline for a system as a movie
recording of past history, using a panoramic lens.  This throws us a
number of questions about how often one should sample data. 

\subsection{Sampling rate}

The naive view in the industry is that one should collect as often as
possible.  Basic information theory constrains our ability to extract
information from data. Many engineers feel that the virtues of fast
sampling are indisputable, just as the citing of many decimal places
leads to increased accuracy, but neither are true (for the same
reason).  Excessive use of high resolution sampling is a senseless
arms race (a watched pot that never boils).  Continuous high density
sampling of a non-existent signal is not helpful.  Nyquist's theorem
tells us that we can only know fully of changes that occur half as
fast as the rate at which we sample.  Shannon's theorem tells us that
our information about the system only increases when changes are
observed.

Regarding the order of events, it's common to rely on an independent
clock service, located within a network to try to synchronize clocks
to some calibrated count, and then rely on the homogeneity of
manufacturing in chip-sets that count time at a more or less similar
rate.  Traditional clocks services count time in seconds, but this
sampling rate is much too infrequent to distinguish processes in
modern processes, where nanosecond timing discriminations are
becoming.

Physics tells us that relying on the counting of an exterior agent is
futile when clocks are located in regions of very different
gravitation, or when they are moving with respect to one another. This
already has to be corrected for in satellite systems. The same effect applies
if agents are in virtual motion with respect to one another. Only the
interior proper time of a process can be relied upon for comparisons.

Local measures of observables have to be aggregated into coarse grains
in order to measure them against one another.  Histograms of
observational distributions are usually the best we can offer in terms
of observability.  But distributions only tells the past probability
of behaviours in fairly static cases.  When we most want to know about
a system, that's when it's hardest to understand.  \footnote{This is
  the paradox of weather forecasting: when nothing is changing, we can
  predict the weather easily; but, when everything is in flux,
  prediction is hopeless.\cite{certainty}.}

\begin{figure}[ht]
\begin{center}
\includegraphics[width=6cm]{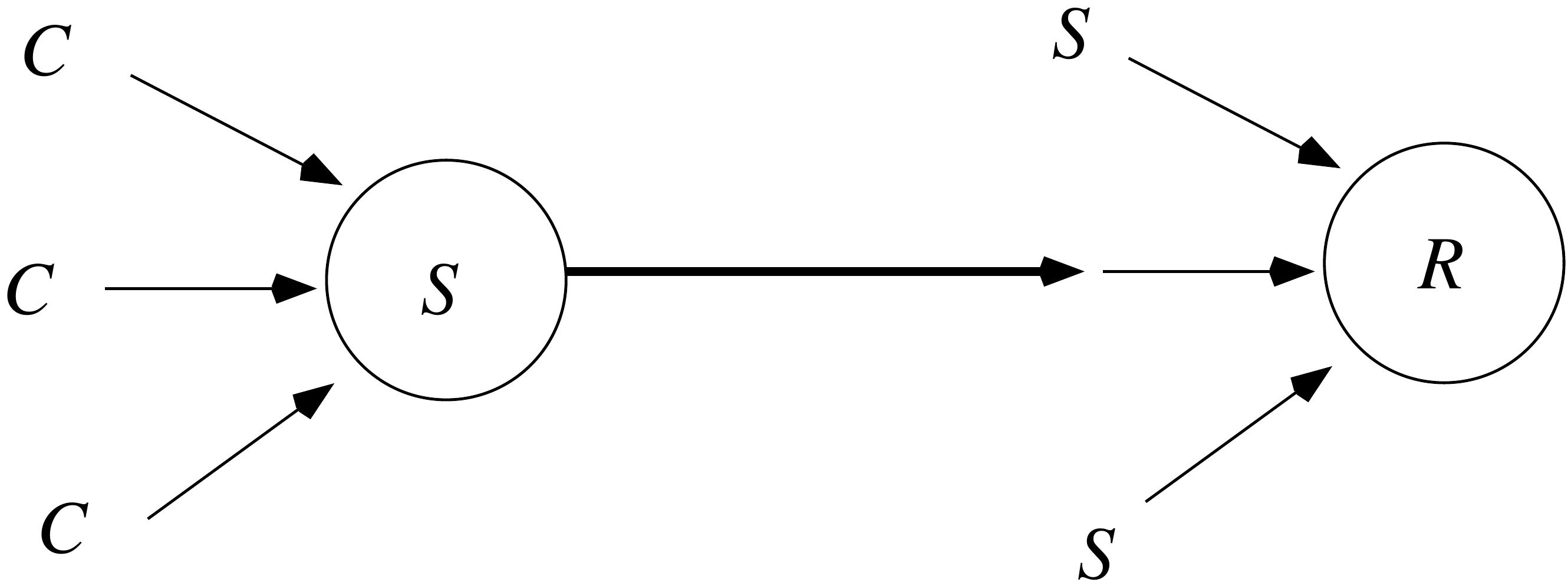}
\caption{\small Aggregation of observations from multiple sources can
  happen at any node in a distributed process.  When causal influences
  come together, in this way, the confluence point becomes an
  effective observer of the sources that feed into it. Observers are not
only human!\label{aggreg}}
\end{center}
\end{figure}

Time-like changes are normally assumed to be instances of what may
potentially be significant events. The result is that human operators
get excited by graphical traces that suddenly rise or fall---which has
an undoubtedly hypnotic appeal, but means nothing without a larger
context.  Sliding windows are often used to detect gradient changes in
time series. Ensemble averages are used and even forced in data
distribution and consensus processes (see figure \ref{aggreg}).
In other words, aggregation over time (not space) is a necessary
part of the learning that provides context for prediction.

\begin{principle}[The sampling rate]
  The sampling rate for a variable should typically be about half the
  auto-correlation time for a variables in order to detect meaningful
  stable variation.
\end{principle}
This is the timescale suitable for learning. For the purpose of
anomaly detection, one might see a sudden change in timescale as a
result of an unexpected coupling.  Faster sampling could then be
introduced on suspicion of a transition in behaviour---just as
biological heart rates and attention spans quicken under stress.
Recording and storing reams of data that are zero or constant cannot be
in anyone's interest. Such data is compressible. It contains no new information.
The potential problem with that approach is that the cost of sampling is not free; the
impact of sampling on the system may become significant. Some authors
have advocated such adaptive sampling\cite{honeycomb}. One then has the decision
about which part of the sampling process to scale back: the act of measuring on each local
process has one cost, the act of aggregating the samples in some central repository
has another cost. Neither of these is easily controllable, since multitasking operating systems
make the sharing decisions to allocate cores, interrupts, network transmissions, and tasks quite opaquely.
It may be difficult to assess which is the greater evil: uncertainty due to 
adaptive sampling or uncertainty due to oversampling or undersampling.

\subsection{Shared resource counters (kernel metrics)}

The consequence of lemma \ref{agglem} is that scaling of observations
causes not only a reduction of information transmission rate, but
possibly a loss of information about the origins of shared
assessments.  Resources counters, computed from the aggregation of
data (like most of the kernel resource metrics typically recorded in
popular monitoring tools), erase details that belong to their higher
resolution origins, in an unrecoverable way (see section
\ref{aggsec}).  There are many such recorded values in timesharing
computer systems, because they are useful mainly to the sharing agent.
The fact that they are shared with separate processes is a mixed and
slightly misleading blessing. It's sad to see so much effort expended
in sharing noise to observers desperate for insight. I think we can do
better.

Shared measures erase the information about data origin, and thus such
collective phenomena cannot be traced backwards to an appointed cause.
For example, measuring the load average for a computer cannot
determine which programs caused a spike in the load\cite{cockcroft1}.
This is not, on the other hand, a reason to retain every individual
data characteristic for ever, because the opposite is also true: some
data cannot be exposed without computing those high level functions.

\subsection{Instrumentation of processes}

Our symbolic representations of processes are algorithms expressed as
program `code'.  In a distributed system, programmers have begun to
instrument workloads for interprocess communication using service
meshes and centralized logging services. These do not reveal
behaviours interior to the processes, but may provide traffic patterns
by which to hypothesize about process behaviours and intentions.

For process tracing, one needs to be in the inside of the process,
where the interior time ticks.  From the foregoing discussion, it's clear that the
system clock service is not the correct measure of time for process
diagnostics. The proper time of a process is measured by the number of
program counter transitions or program steps incurred by its
execution. This count is significant because each increment is the
result of an explicit causal jump instruction.  The program counter's
value can be promised at each instruction in code for debuggers to
trace, forming a causal set of correspondences between program
locations and increments. However, attempting to share these proper
times between processes is meaningless.

The system UTC clock, even an approximate local copy of it, comes from
an independent agent, and although shared between many processes its
increments do not correspond to single channels of causation. Rather
they represent only the implicit increase of entropy in collecting from all
processes.  When processes are timeshared, they may be halted and
interleaved in complex ways.

\subsection{Significant Events}

As discussed above, the major drawback in time-series thinking, for a
distributed system, is that there is no unique meaning to the order of
transactions originating from different sources when they are
aggregated from different locations.  Each observer in the universe
sees events from their own perspective.  The lightning bolt and the
thunder arrive earlier for some than for others, because different
processes propagate information at different rates, over different
routes, and with different latent delays.  Consensus is expensive,
heavy handed, and its goals are different to the goals of observation.

Moreover, we have no uniform metric for time other than the exterior
clock, which has issues of its own---it doesn't represent local
causation except for the process that generates it. The question we
want to answer is: what was the reason for an event $E$? i.e.  can I
infer the condition, state, or quality of the system from this event, based on
what I have observed beforehand?  Times are not causes.

A better approach, based in interior instrumentation is to create a
semantic counter that traces a distributed process that we can trace
backwards to causal origins. Samples can be taken when something is
found to be significant within the context of the process itself. This
is the proper passage of {\em significant} times.  Regular sampling of
processes is not an efficient way to record them because processes may
be busy or idle, etc. This is what system logging enables---but the
opportunity is usually squandered from a lack of a proper model.

Metric coordinates (clock times and numbered locations) are not
helpful when we have no invariant measuring devices to define them by.
The alternative is to use descriptive labels, or semantic coordinates.
\begin{definition}[Significant event]
An event marker, provided by a source process, that signals
either an intentional change or an unintended deviation from
expected state.
\end{definition}
Anomalies and faults fall into this. Processes that are not
able to keep their promises may also be significant events.

When seeking the `root cause' of an event, we really want to go back
to prior events that were significant.  The rationale behind this is
that, in a stable system (one that we expect to be predictable), when
all is as expected, unexpected events are most likely to be caused by
previous significant changes and anomalies.  Clearly, for every event
there is a prior one, until we reach the very beginning of time (the
system `Big Bang'). However, we also have episodic boundary conditions
that act as `Little Bangs' for more constrained universes.  Boundary
conditions are semantically special events that we attach special
significance to---they are the origins of causation. We aren't
interested in every intermediate change, only in prior events that
make a splash.

We want to create a causal chain, a journal, something like a linked list.
Instead of selecting data by voting, we can select values based on their
perceived importance to outcomes of interest. This shifts the focus of
policy from the intermediate aggregator of data to the observer: the observer
is now expected to have a specific question it wants answered, rather than
voyeuristically consuming data for entertainment.

\subsection{From metric to semantic significance}

The use of `signposts' for labelling paths, as semantic
coordinates\cite{smartspacetime}, traces back before maps and
calendars were invented by human civilizations. Instead of imagining
idealized coordinate grid systems, perhaps unmeasurable, signposts
relate events to less regular but highly recognizable things that we
can observe (the big tree at the river, Mount Fuji, the Matterhorn,
the year of the flood, the eclipse of the moon, etc).  This provides
anchors for accessing memory and plausible anomalies that might have
exerted causal influence.  We use these events as boundary conditions
on episodic sub-processes.

Descriptive naming (semantic coordinate assignment) is thus more useful than
ordinal naming (numerical coordinates).  The checkpoints and paths
that participate in processes may not be invariants, and the numerical
value of coordinates is irrelevant\footnote{One works hard to make this
  point in the physics of relativity where coordinate systems get in
  our way of understanding spacetime processes}. We may need to
identify a repeated pattern to some degree of approximation in order
to exemplify a general concept from which lasting knowledge can be
derived.  Anomalies that do not recur become effective invariants, in
memory, because they are rare and worth remembering. Featureless
invariants (like empty space) are indeed the most invariant of all,
but have such high entropy as to contain no information of
significance.

The {\em significance} or meaning of a signal is the heuristic inverse
of the (incompressible) information within it. The more information
we need to characterize a room, the less stands out about it. If there
is one part that dominates, the rest is negligible---hence the
principle of signalling significance.
\begin{lemma}[Significance vs information]
  Maximum entropy distributions contain no significant events: they
  are causally random, and all events are observationally equivalent.
  Minimum entropy distributions have the maximum significance, as they
  imply strong correlation.
\end{lemma}
Entropy plays a subtle role in statistical distributions,
and therefore in ability to infer meaning from data.

\subsection{Reversibility versus traceability}

We want to be able to trace our knowledge of a system back to know the
cause of an effect. The intended outcome is programmed into it, but
there are also unintended outcomes caused by environment leaking into
causal pathways. Because of the culture of `rollback' thinking in IT,
which originates from database transaction semantics, IT often muddles
the concept of traceability with reversibility\footnote{Reversibility
  is something of a misunderstood concept in dynamics, especially as
  applied to IT system behaviours. The apparent reversibility of the
  machinery is sometimes used as an argument against causality, but
  the argument is built on a misunderstanding that ignores the
  boundary conditions.}.

We must distinguish between the ability for an observer to trace
backwards from a sequence of observations to reconstruct the cause of
a significant event, and the ability to roll the state of a system
back to what it once was. For example, it's possible for an observer
to trace the source(s) of a river, but it is not possible to reverse
the river and roll it back to an earlier state.

In the former case, the enabling condition is for no origin
information to get lost in the chain of unfolding events (see figure
\ref{order}).  In the latter case, the necessary condition for being
able to undo a causal sequence is that agents of the system itself
have to promise the inverse of every promise in the forward direction
conditionally on an undo condition---this is an additional set of
promises pointing backwards along the path, which is much more than an
observer being able to trace knowledge of promises backwards. The
necessary condition is insufficient even to promise the result: agent
processes must also be isolated from external interference, else the
precise inverse operations may be deflected off course by
noise\cite{jan60}.

\subsection{Partial order of agents and events}

As a process propagates by passing messages, the messages separate
earlier times from later times on the process's own clock or counter.
Suppose this is reflected in a sequence of messages, or lines in a log.
In a chain of lines $L_\gamma$ belonging to a single source $S_i$, 
we can define a countable metric distance between lines by the total ordering
of the sequence, also in the language of promises:
\beq
L_1 \promise{+\text{precedes}} L_2  \promise{+\text{precedes}} \ldots L_n \promise{+\text{precedes}} L_{n+1}.
\eeq
The observational binding is equivalent a more classical ordering relation $<$:
\beq
L_1 < L_2  < \ldots L_n < L_{n+1}.
\eeq
where
\beq
S < R ~~~~ \leftrightarrow ~~~~ \left\{     
\begin{array}{c}
S \promise{+\text{precedes}} R\\
R \promise{-\text{precedes}} S
\end{array}
\right.
\eeq
We need to be cautious about the interpretation of these promises.
Each agent is making a separate promise, but (by the law of agent autonomy)
these agents cannot make the assessment or promise it to an observer who happens
to be watching all of them. Each agent can make its own promise available to
the observer, but it's up to the observer to order them in the final instance.
This ordering may be come mixed with other orderings as data are aggregated.
\begin{figure}[ht]
\begin{center}
\includegraphics[width=7cm]{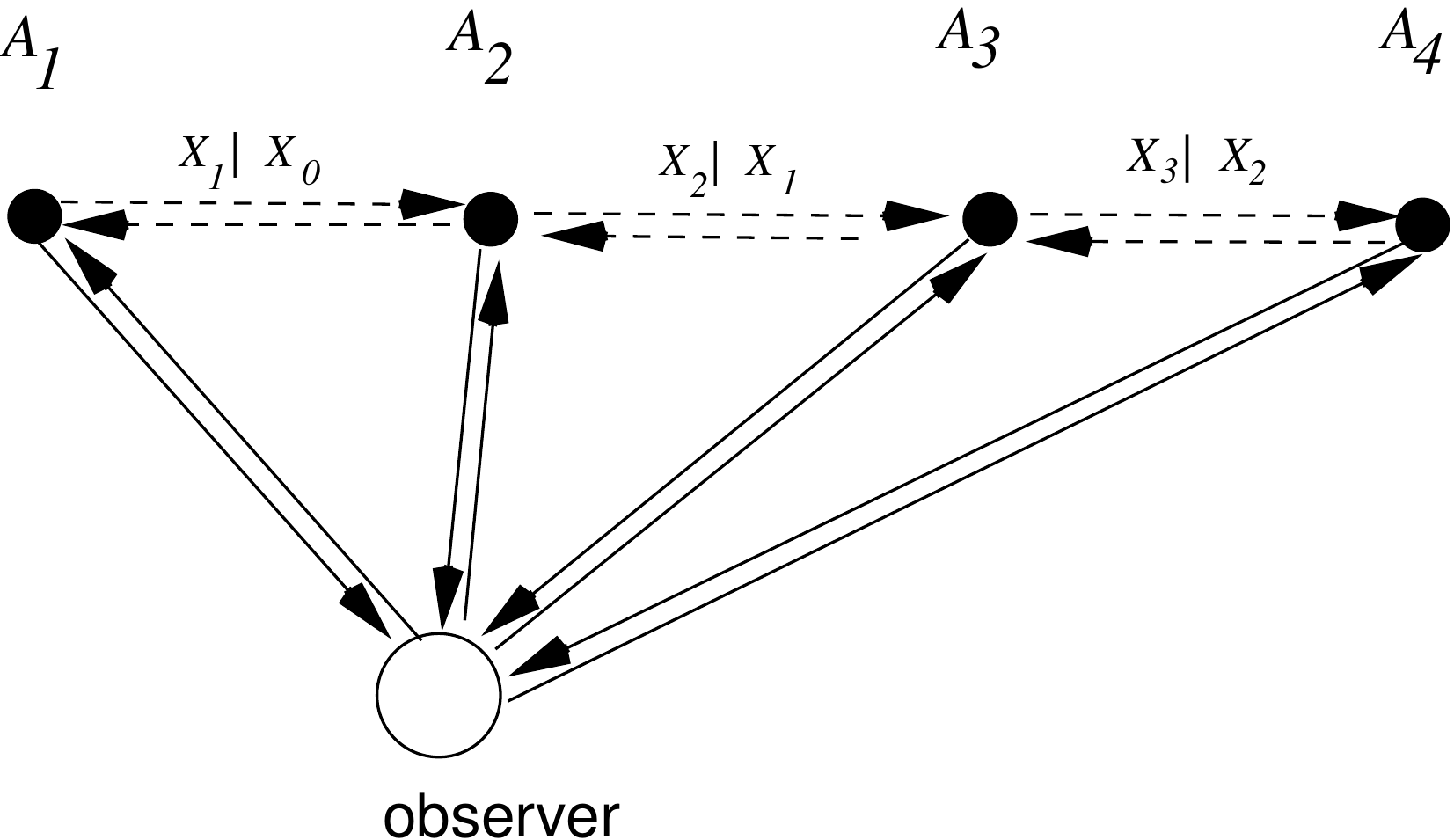}
\caption{\small Causally ordered change in a process, and information observed
about the process are two distinct things. As long as the observation of the process
retains the order of the process, inferences about causality can be made, regardless
of whether the system itself could be reversed. You can trace the source of the Nile,
but you can't make the river flow backwards.
  \label{order}}
\end{center}
\end{figure}
The promises indicate that the relationships are considered persistent
or even invariant by the promisers---not merely local assessments made on the basis
of spurious data. However, if at any time the events being ordered become
indistinguishable, they can no longer be ordered. This can happen when data are
aggregated without complete labelling.

Ordering reduces to the existence of (local) conditional promises,
whose scope may extend to other observers in a scope $\sigma$\cite{promisebook}.
The $n$-th agent in a sequence; by the axioms of Promise Theory, we must have
a chain of the form (figure \ref{order}):
\beq
A_{n} &\promise{+X_n | X_{n-1}}& A_{n+1}\\
A_n &\promise{-X_{n-1}}& A_{n+1}\\
A_{n+1} &\promise{-X_{n}}& A_n\\
\eeq
And, in general, we may consider a general scope for the above promises:
\beq
A_{n} &\scopepromise{+X_n | X_{n-1}}{+\sigma}& A_{n+1}
\eeq
Given such a set of promises, we can define a measure of observable distance
between agents $A_n$ and $A_m$ by assessment.
Again, we note that interior relativity makes distance 
an assessment by one agent about the relationship between itself and one or two others.

\subsection{Translation operator and Noether's theorem}

If the agents were sufficiently homogeneous, we could consider an
operator interpretation for $\Delta$, as the generator of a translation on a set of
states realized by the positions (like ladder operations):
\beq
\Delta | A_i\rangle \rightarrow | A_{i+1}\rangle.
\eeq
In fact, for every kind of promise there would be a separate propagator, like a complete basis:
\beq
\Delta = \sum_\tau \; c_\tau \,\Delta_\tau.
\eeq
The problem with this kind of interpretation is that is suggests the existence of
a god's eye view once again. It takes the existence of a privileged observer to be able
to order and rank the states in this way.

In classical physics, the continuity of the energy function with
respect to spacetime is what generates conserved quantities like
energy and momentum, thus allowing these quantities to be used
consistently as counters for behavioural descriptions.  We can see,
from the Promise Theory, that this conclusion also follows from a
privileged god's eye view of spacetime locations.

This tells us that it is the {\em assumption of continuity} by the observer
that rationalizes the use of counting metrics, including jumps and changes
in metric behaviour. If source observability does not reveal discontinuities
in the assumptions amongst independent sources, the observer will not be able
to discern that information merely by monitoring.

If we assume the conservation of $b$ as an axiom, the ordering of
$b$-influence must follow paths automatically, even when the agents
make unsynchronized (asynchronous) promises, like a first order Markov
process. In order to {\em explain} conservation and causal order over
non-local regions, we need to extend the promises to be conditional on
non-local neighbouring patches.  Ordering information itself needs to
propagate.

\subsection{Causality}

The principle of causality can be stated simply by saying that earlier
events are followed by later events, at a given point of action, as a
result of the transmission of some information that we may call an
influence.
\begin{definition}[Causality]
A causal graph is a complete graph of conditional promises.
We say that $S$ causes influences $X$ at $R$ with cause $c$ iff:
\beq
S &\promise{+X_S\;|\;c}& R\\
R &\promise{-X_R}& S,
\eeq
and $X_S \subset X_R$.
\end{definition}
If every transfer of influence in a system obeys this
property, then we can define the system to be reversible:

\subsection{Traceability (inference)}

\begin{lemma}[Traceability]
If an observer has complete information about promise causality, a process graph may
be called reversible, i.e. for every pair
\beq
S &\promise{+X_S\;|\;c}& R\\
R &\promise{-X_R}& S,
\eeq
provided $X_S \subseteq X_R$. 
We can infer origin by using complementarity to
interpret a reversal of causal tracing: $\overline X_R = c$ and $\overline c = X_S$, such that
\beq
R &\promise{+\overline X_S\;|\;\overline c}& S\\
S &\promise{-\overline X_R}& R.
\eeq
\end{lemma}

If a set of agents $A_n$ precedes another set $A_{n+1}$ by a promise
\beq
A_n    & \promise{+X_n | X_{n-1}}& A_{n+1}\\
A_{n+1}& \promise{- X_n}& A_n.
\eeq
Traceability requires that $O$ be in the scope of this chain, and that it assumes
reversible semantics for $X$, as `is followed by' (which is automatically
interpretable as `follows').

\subsection{Reversibility (causation)}

This may not point out a unique `root' cause, but it will point out the causal
sets that act as source (spacelike hypersurfaces) of the process.

\begin{lemma}[Reversibility]
 requires, much stronger:

\beq
A_{n}    & \promise{+ \text{Inv}(X_n) | \text{Inv}(X_{n-1})}& A_{n-1}\\
A_{n-1}& \promise{-\text{Inv}(X_n)}& A_n.
\eeq
This holds for any agent (or superagent) $A_n$.
\end{lemma}

The condition for forensic back-tracing of a system state (detection of cause)
is that 
\begin{itemize}
\item A complete chain of prior origin data be available across the
graph. 
\item There should be no acausal loops in the process,
else there may be branch alternatives (eigenstates) or
divergent unstable behaviour.
\end{itemize}

\begin{example}[Service lookup thunder and lightning]
Consider the order of a process (figure \ref{causalorder}) described in the
following promises:
\beq
A &\promise{-\text{dns}}& S\\
S &\promise{+\text{dns}}& A\\
A &\promise{+\text{relay(dns)}|\text{dns}}& R\label{cond}\\
R &\promise{-\text{relay(dns)}}& A
\eeq
\begin{figure}[ht]
\begin{center}
\includegraphics[width=7cm]{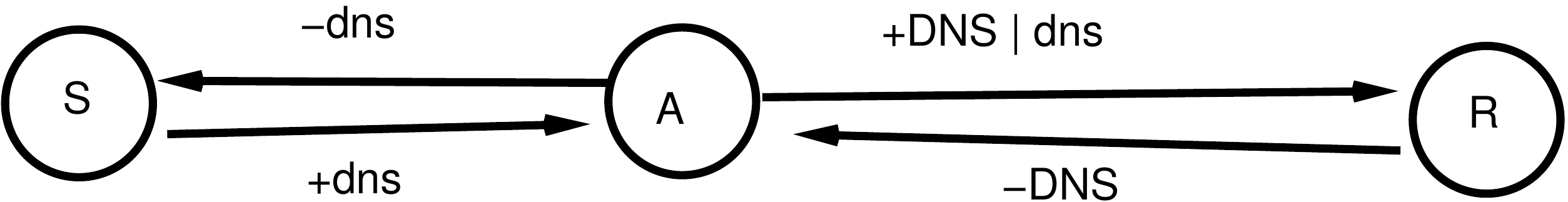}
\caption{\small Causal order may be different from clock time. It is
  generated by prerequisite dependencies: either by underlying
  topology or by constraint. Agents can only trust directly agents
  that they are in scope of (in practice, their direct neighbours), as
  they have no calibrated information about the promises of agents.
  \label{causalorder}}
\end{center}
\end{figure}

Agent $A$ promises to listen for a DNS lookup (a query, i.e.
an invitation to reply). As long as this promise exists, it can be considered
to be polling $S$ for a response. $S$ promises to provide DNS data, but it hasn't
specified what or when. If there is a fortuitous match between the two, data
will be passed from $S$ to $A$. $A$, in turn, promises $R$ to pass on the data
it receives from $A$. It does a better job of promising conditionally, so it will
only pass on fresh data from $S$ when a reply is received, because the promise
in (\ref{cond}) is conditional. $R$, in turn, promises to accept data, which
can only happen after they arrive. The effect in each interaction is to order
data, but we don't know, from these promises, how many times data get passed
between $A$ and $S$, nor do we know how much latency is experienced by any of the agents.
\end{example}

The conditional promises (dependency) represent causal ordering.  We can't say
anything about the relative order of promise keeping unless it is
constrained in some fashion. Often we rely on incidental or ad hoc
serialization at a single observation point (a queue) to define the
ticks of our process clock. The problem is that this serialization
does not represent an invariant of the process, so it's unreliable.

\subsection{Metric distances}

Order is important, when it can be distinguished because it allows us
to measure intervals. We sometimes use intervals as significant
measurements, though Einstein pointed out that intervals are not
invariants, they are only `covariant', changing with the system of
measures we establish.

\begin{principle}[Distance semantics]
Distance is an assessment made by an observer with two complementary
interpretations: distance suggests what might lie in between the bounds
of the interval, or it suggests a measure of how similar two agents
are, with respect to location in some criterion `space'.  
\end{principle}
The distance between two events $L_\gamma$ and $L_{\gamma+\beta}$ is
related to the ability of an observer to trace and count the number of
similar events in between. The distance between events in a journal
may contain implicit information about what happens in between, but it
is not a substitute for the information itself. Metric distance is
therefore a counter that pays just enough attention to agent
properties to discriminate between them on the basis of label, and be
able to count, but not necessarily enough to classify agents
meaningfully.  If events follow on as nearest neighbours this tells us
something; if the same pattern is suddenly interleaved by more lines
this could be an indication of an anomaly.

A histogram is a classification of multiple events that get counted
and form a distribution. The order of the classes may (or may not)
express a metric policy about how near or far events are when they
fall into one of the classes. There is no a priori order to these
classes, but there might be a distance.  It's therefore an assessment policy of an observer to ensure
proper classification according to a model presumed by the observer.

The proliferation of logs in IT systems means that they get receive
disproportionate focus, in the hope of extracting far more than they
are usually capable of representing. The variety and standard of
logging is very poor indeed, in my view.  What happens to order and
distance relationships in logs after aggregation?  There are many
tools that imply log aggregation is a good way to bring together all
logs into one location, but there is little discussion around the
significance or usefulness of the result\cite{stadler4,matt,gossip}.

Aggregation of agents $L_\gamma$ into superagents $\{ L_\gamma \}$ may
preserve or discard the order and interval distances between lines.
Sometimes, data are not intentionally numbered by the sender and order
is assumed by the order to transmission (e.g. in UDP transmissions).
In that case, message packets may become reordered by network
redirection, or loss.  Some messages could be also be lost. Let's
refer to the cases by the common terminology
\begin{itemize}
\item Reliable: promises all packets delivered in order.
\item Unreliable: ad hoc, no promises about order or loss.
\end{itemize}
In either case the latency between transmission and final arrival is uncertain.
Consensus of data is easy, because the data are point to point and there is
only a single source and a single receiver for each message.

One way of trying to work around the law of intermediate agents is to
build up the notion of {\em entanglement} between processes \cite{paulentangle}.
This takes several cycles of mutual interaction between a pair of agents,
on some scale, as well as a small cache of local memory.
Entanglement can transform partially reliable transmission of influence
into fully reliable transmission at the expense of some added sub-cycles
in the interaction.

The aggregation of messages without reference to the agent and the
interior timeline that generated them implies that causal origins
can never be traced backwards.  Timestamps have no value, because they
are unrelated to the process causation.  We can thus show that a log
may preserve the reverse tracing of causal history, but does not imply
reversibility of state. We can trace a story back to where events
played significant roles in the timelines of processes, but we can't
necessarily reconstruct the states of those processes.

\section{Aggregation of source data}\label{aggsec}

It's time to look more carefully about what aggregation of data means.
In the context of causality\cite{stadler4}, it matters both where and when signals
come together, and to what degree information is lost by mixing.
Distinguishability plays, again, a central role here.  For example,
the commonly used metrics of load average, CPU percentage, memory
usage etc.  The behaviour of a process depends on the behaviour of the
platform, which in turn depends on the behaviours of the guest
processes\cite{cockcroft1}.  These are measures of different scales, since a platform
is an aggregation of processes, and so on.

When unexpected behaviour (signpost behaviour) is observed in an
aggregate variable, the culprit may not even be in the same process.
The relevant question may seem to be: can we obtain information about
which process may have been responsible? But that is not the right
question, because it could be the accumulation of many processes
leading to an exhaustion of resources which actually impacts the
process we are monitoring---by undermining its critical dependencies.
When this is the case, we might be more interested in why scheduling
policies resulted in such a confluence of demand. Obviously, there are
many layers of decision behind such stress concentrations.

The causal connection between these cannot be inferred with any
certainty from quantitative measurement however. One would rather
expect to see a process log fail to allocate memory from within the
privileged context of the process itself. Today, we wrap processes in
containers that are quite opaque. In fact, processes are equally
opaque when viewed through kernel metrics, because there is
unfortunately little or no causal connection between the changes and
any particular process of interest.

\subsection{Sampling resolution (timescales again)}

We need to know when data belong together and when they should be
considered separate. For any collector, this is a policy decision, but
it can be informed by the physics of the system. There is information
in the order, content, and volume of data. If that information is
squandered, it may be unrecoverable.

\subsection{Erosion of metric significance}

Experiments show that there is little correlation between commonly
collected quantitative metrics and actual process
semantics\cite{burgessC8}. This is a historical artifact that comes
from the fact that observables were designed for timesharing, not for
process monitoring.  Measuring kernel metrics is something analogous
to watching the weather to plan for a crop. In some cases, a change in a distant place
may may trigger outcomes that result from arbitrary choices in code
elsewhere. For example, if one sets an arbitrary threshold for a
value, in a conditional statement, the unusual process weather
originated elsewhere may push the conditional over the limit
unexpectedly and lead to a discontinuous branch of behaviour.  This is
why understanding relativity is so important in reasoning, and why
cloud computing is especially susceptible to relativistic effects.

Entropy of mixing does not usually increase relentlessly in IT systems, because
new information is being added in the form of semantic labels (boundary conditions)
all the time, e.g. when a particular set of images is identified as 
belonging to the same person, we name the set as the person; or when
a sequence of command instructions leads to the same failure mode, we name
the histories with the name of the failure mode. This new information
adds context.

As data scale, some information is lost, and new information is added.
Aggregating data and integrating over time, throwing away time information,
but building maps of invariant relationships. The map of what remains
distinguishable grows as more data
are added, because the number of possible storylines grows as new invariants
are added.
\begin{lemma}[New data at all scales]
  Origin data are lost by coarse graining, but combinatoric selections
  of aggregates leads to a new degree of freedom: distinguishable
  routes or paths through the composite variables.
\end{lemma}
Each story has its own semantics: the loss of event indexing
leads to the addition of fewer semantically stable storylines.
Distances that require distinct labelling become meaningless,
but new emergent distinctions lead to new possibilities for
classification.
If sufficient information is retained to point backwards along
to causal signposts, specific paths can be traced without
muddying the global picture.

\subsection{Learning and coarse graining defined}

We can track the different scales of a system by seeking to separate
invariants from microscopic local changes, using the principle of
separation of timescales. Learning over sequences (in a timelike
direction) effectively form Bayesian processes that can act
as aggregate state discriminators\cite{pearl1,pearl2}.

\begin{example}[Learning]
  The collection of data to train a statistical algorithm may take
  weeks or months, and involve large amounts of data that are
  compressed into a composite form, irreversibly. The composite
  aggregate is used as part of an algorithm that recognizes images on
  a timescale of seconds. These processes can be naturally decoupled.
\end{example}
The degree of separation of timescales corresponds to what we call
supervised learning (highly separated timescales for learning and
using) and unsupervised learning (where timescales for learning and
using are approximately the same).

Even a single unique episode may eventually become viewed as an invariant
if it is not repeated and hence is never challenged, but its
significance may be limited. We can only know this by learning over
time. The significance of concepts grows by the frequency by which
they become repeated. Thus learning and garbage collection of
insignificant concepts is needed to prevent all the information from
becoming noise.

For full episodic reconstruction, the invariant connections that
generate process stories need to integrate with one another, like a
linked list, using the a map of invariant concepts as glue. The
invariants represent the aspects of processes that are not specific to
a single source. See the earlier work on characterizing spacetime
semantics \cite{spacetime1,spacetime2,spacetime3,cognitive}, based on
earlier experiments \cite{stories}.

All causality is a representation of Shannon's basic model of an
information channel.  The distinguishability of information is the key
to following and tracing processes, but where does the significance of
information lie?  The significance of information (associated with
labels to it) is necessarily diluted by scale, or the entropy of
signal aggregation.

\subsection{The Mashed Potato Theorem}

Mixing of signals leads to loss of traceability.
Suppose you are at a restaurant and you receive some mashed
vegetables.  You first assume that it's potato, because that is
common, but something doesn't taste quite right. There are some other
vegetables mixed in.  Closer inspection reveals some orangy colour
(perhaps carrots, or sweet potato, etc). How could you know what was in the
potato without accurate knowledge from the source?

The loss of distinguishability (entropy of mixing) tells us that we
can't easily discern the content of mashed potato without the recipe
because we cannot separate (classify) the parts of the signal.

\begin{theorem}[Loss of distinguishability]
Let $\Sigma$ be an alphabet of class categories that are distinguishable
by a set of source agents $S_i$. 
Data aggregated from $S_i$ without complete causal labelling $\sigma \in \Sigma$, from the source
cannot be separated into its original categories $C_\sigma$ with certainty, i.e.
the promise
\beq
\pi_\text{categ} : R \promise{+C_\sigma|\text{data}_\intersect} \Unspec\label{xx}
\eeq
is kept with equal probability for all $C_\sigma$.
\end{theorem}
To see this, we note that the aggregation of data involves promises:
\beq
\pi_\text{data}: S &\promise{+\text{data}_S}& R\label{xxx}\\
\pi_\text{listen}: R &\promise{-\text{data}_R}& S\\
\text{data}_\intersect &=& \text{data}_S ~~ \intersect~~ \text{data}_R.
\eeq
and we take data to be a collection of line signals $\text{data} \sim \{ L_\gamma\}$.
In order for the conditional promise (\ref{xx}) to be causal, the promise
of data in (\ref{xxx}) have to be a reversible function of the $C_\sigma$. But if
data are indistinguishable, then the $C_\sigma$ must also be indistinguishable, thus
\beq
C_\sigma = C_\tau, ~~~\forall \sigma,\tau
\eeq
thus, the probability of discerning a signal $C_\sigma$, $p_\sigma \equiv P(C_\sigma)=p_\sigma$, 
and the entropy
\beq
S_\text{ent} = -\sum_\sigma  \; p_\sigma\; \log p_\sigma \rightarrow \max({S_\text{ent}}).
\eeq
If one rescales all the $C_\sigma=C_\tau$ into a single category to
indicate that all such categories are the same, then the entropy is
zero, $S_\text{ent} = 0$, indicating that the information per
transmission, in the mixture, is actually trivial. Thus, we must keep
all labels from different sources and classifiers in order to retain
useful information. This does not depend on the amount of data (or
mashed potato). Moreover, if data are passed on, the dependence of a
reconstruction by (machine) learning is unstable to the
datasets\footnote{I won't consider this here, but see the signals
  lemma in \cite{william2}.}. The definition of entropy in computer processes has been examined
recently to address its semantics\cite{algentropy}.

\subsection{Separation of concerns}

The practical question remains: how should one separate variant and
invariant data when designing systems?  This depends partly on the
structure of intent and observations.  Programmers are trained to
recognize what values are variable and abstract them into parameters
to invariant functions.

If we think about how we formulate stories, as humans, we embed
variable fragments of causal history (episodes) into larger assemblies
of more invariant concepts, which provide context for reasoning. This
is how experiences are organized around conceptual models. I'll come back to
this in section \ref{model} on model extraction.

\begin{example}[Logging text compression]
As a simple example, consider the generation of a log message from a
typical format string in code:
\begin{itemize}
\item A separate format string is an invariant class of messages.
It can be replaced by a single numerical value and looked up in a hash table
to compress data.
\item Standard data format can record format string and variables in an 
indexed structure with named members.
\item Message significance level or priority (policy) - imposed + or -?

\item Variable Substitutions in the format string are variants with respect
to the message. Some of these may be invariants too (the name of a host or function),
while other data have no long term significance (the time or date).
\end{itemize}
If we compare these points to a Unix syslog message, the glog library,
and many more examples, it's clear that syslog satisfies none of these
promises. Lines of text are basically random.
\end{example}

\subsection{Retaining semantic context for events}

Concepts are the result of dimensional reduction over contextual
learning sets $C_i$ from a number of sources $S_i$ \cite{spacetime3}.
In invariant cases, the context can be learnt by accumulation of
evidence over time, because it doesn't change. In general,
significance may be assessed based on a number of contextual sets
$C_i$, so when an alert message $M$ is reported to an agent $R$, this
is in fact a conditional promise that depends on the context:
\beq
 S \promise{+M|C_1,\ldots, C_\sigma ~\forall i} R
\eeq
This means, by the conditional promise law, that the promises supplying this
context to $S$:
\beq
A_i &\promise{+C_i}& S\\
S &\promise{-C_i}& A_i,\\
S &\promise{-C_i}& R,
\eeq
are not available to $R$. The context is lost.
This means $R$ has to trust the alert and its significance as a random variable.
This is no problem if the goal is to bring an unrecognized condition to the attention of
an operator. However, if the goal is to perform contextualized reasoning
on an aggregate scale, the graph of invariant context also needs to be promised:
\beq
A_i &\promise{+C_i}& R\\
R &\promise{-C_i}& A_i,
\eeq
and the conditional dependencies also need to be captured:
\beq
 R \promise{- (M|C_i, ~\forall i)} S
\eeq
If we didn't apply this idempotently only to sparsely occurring invariant concepts, 
the cost of the aggregation would rise sharply. An expedient 
separation of scales allows the context to be contained at the sources
as `smart sensors' \cite{spacetime3,cognitive}.

Context can be framed and localized using namespaces.  Namespaces also
provide unifying labels that can
usually be treated as invariants in information systems.
Aggregation of messages, without cataloging, indexing, or other
labelling leads to ensemble entropy: the irreversible loss of
structural information and contextual semantics.

\begin{figure}[ht]
\begin{center}
\includegraphics[width=4cm]{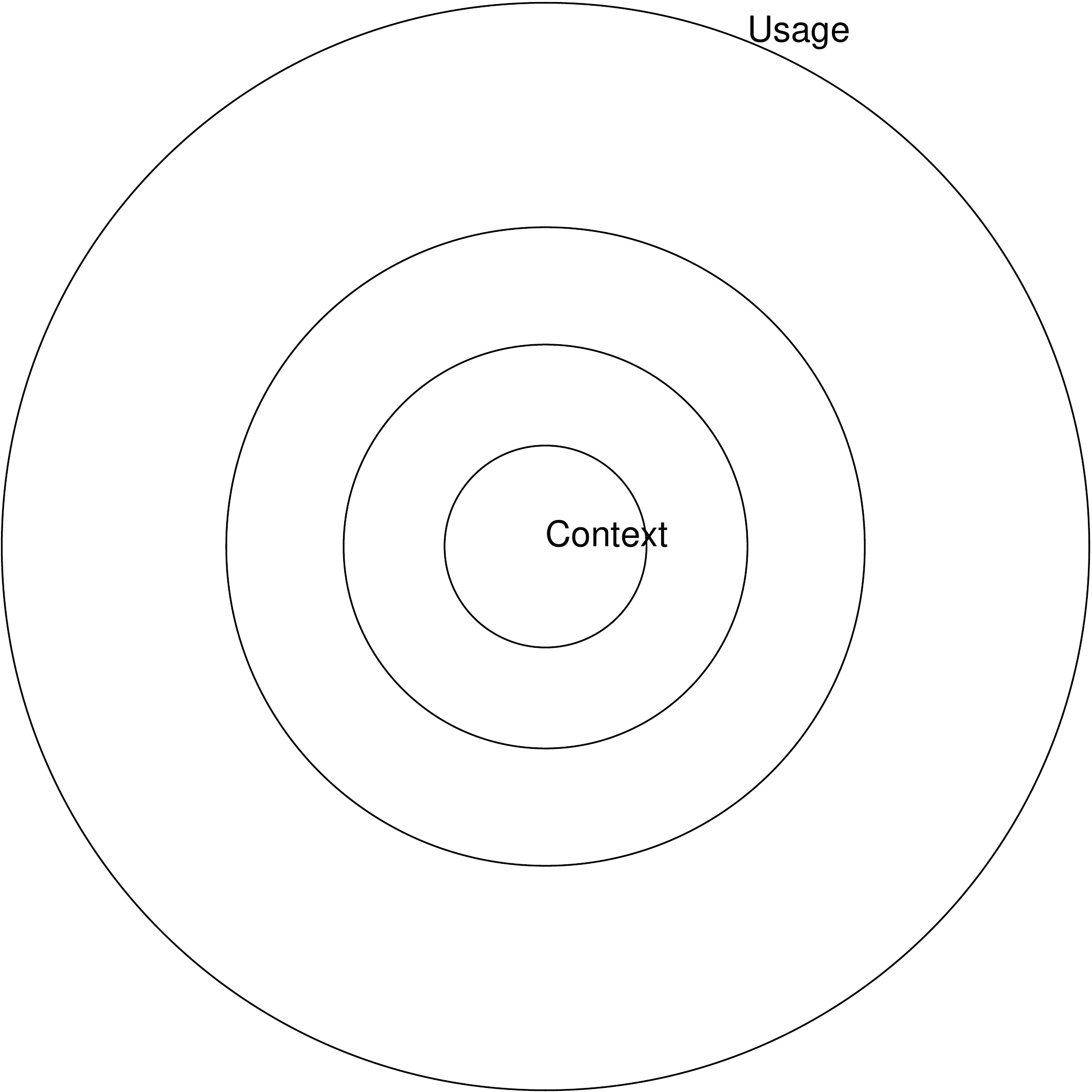}
\caption{\small As data get propagated farther from their initial
  context, their original meaning is degraded, unless all context is
  transported with them.  Each of the rings may represent an
  intermediate agent that may or may only promise to forward data
  selectively or after distortion.\label{usage}}
\end{center}
\end{figure}

Transporting too much context is a questionable idea. If the
environment in which context originates is lost, then the meaning of
the context is also lost, and the ability to reconstruct scenarios
based on it becomes of largely forensic interest. System designers
need to find expedient ways to compress context and filter it: what
can remain local at the source, and what can be aggregated and
assigned wider meaning? Thus is remains a policy decision to balance
the cost of preservation against the actionable usefulness of doing
so.

\section{Histories: logs and journals}

Now aware of the issues around sequentialism and observability in
distributed systems, we can tackle the first two story types in
section \ref{storytypes}.  Logging of process conditions may well be
the most popular and common approach to tracing in computer programs.
Isolated single-agent logs are simple serial queues, or time-series
databases, of varying degrees of sophistication, for keeping
informative messages about what transpires in a process. This is no
longer true for log aggregation unless complete causal linkage
is preserved.

\subsection{Causal linkage}

In most shared logging services, messages are imposed by multiple process agents $S_i$
onto a queue and are strongly ordered by a single receiver $R$.
\beq 
S_i \imposition{+L} R.  
\eeq 
$R$ accepts requests indiscriminately
\beq 
R \promise{-L} S_i.
\eeq 
Individual processes can voluntarily write their own logs but this is not a common practice
because the end goal of logging, in modern practice, is to aggregate
all messages as `big data' to be trawled.

Modern logging services, like Prometheus etc, provide more nuanced
semantics with structured data formats that can incorporate
key-values; but these trust data to be useful. They are abused greatly
by programmers, who tend to dump any and all data into a stream
without regard for meaning or consequence, in the hope of sorting it
all out later.  Logs need to promise invariant causation:
\begin{itemize}
\item Same text (signal) as the same interpretation.
\item Information is encapsulated as transactions to show partial order.
\item Every significant transaction needs to point to its previous significant event.
\end{itemize}
These principles have been embodied in a proof of concept implementation\cite{koaljahistory}.

\begin{figure*}[ht]
\begin{center}
\begin{alltt}
\tiny
New process timeline for ( myApp_name21.2.3 ) originally started as pid  17778 

Unix clock context              | root --> NOW,delta  Comment indented by subtime
------------------------------------------------------------------------------------------
2019-06-03 13:40:04 +0200 CEST  |    0 -->   1,1      MainLoop start 
2019-06-03 13:40:04 +0200 CEST  |       ->   1,2        [function: main] 
2019-06-03 13:40:04 +0200 CEST  |    1 -->   2,1      Beginning of test code 
2019-06-03 13:40:04 +0200 CEST  |       ->   2,2        [remarked: : Start process] 
2019-06-03 13:40:04 +0200 CEST  |       ->   2,3          [go package: cellibrium] 
2019-06-03 13:40:04 +0200 CEST  |       ->   2,4            [btw: example code] 
2019-06-03 13:40:04 +0200 CEST  |       ->   2,5              [remarked: : look up a name] 
2019-06-03 13:40:04 +0200 CEST  |    2 -->   3,1      code signpost X 
2019-06-03 13:40:04 +0200 CEST  |       ->   3,2        [intent: : open file X] 
2019-06-03 13:40:04 +0200 CEST  |       ->   3,3          [file: /etc/passed] 
2019-06-03 13:40:04 +0200 CEST  |       ->   3,4            [dns lookup: 123.456.789.123] 
2019-06-03 13:40:04 +0200 CEST  |       ->   3,5              [btw: xxx] 
2019-06-03 13:40:04 +0200 CEST  |       ->   3,6                [coroutine: main] 
2019-06-03 13:40:04 +0200 CEST  |    3 -->   4,1      Run ps command 
2019-06-03 13:40:04 +0200 CEST  |    3 go>   5,1      TEST1--------- 
2019-06-03 13:40:04 +0200 CEST  |       ->   4,2        [btw: /bin/ps -eo user,pcpu,pmem,vsz,stime,etime,time,args] 
2019-06-03 13:40:04 +0200 CEST  |       ->   5,2        [btw: Testing suite 1] 
2019-06-03 13:40:04 +0200 CEST  |       ->   5,3          [intent: : read whole file of data] 
2019-06-03 13:40:04 +0200 CEST  |       ->   5,4            [file: file://URI] 
2019-06-03 13:40:04 +0200 CEST  |       ->   5,3          [remarked: : file read failed] 
2019-06-03 13:40:04 +0200 CEST  |       ->   5,4            [system error message: open file://URI: no such file or directory] 
2019-06-03 13:40:04 +0200 CEST  |       ->   4,3          [remarked: : Finished ps command] 
2019-06-03 13:40:04 +0200 CEST  |    3 go>   6,1      Commence testing 
2019-06-03 13:40:04 +0200 CEST  |       ->   6,2        [remarked: : Possibly anomalous CPU spike for this virtual CPU] 
2019-06-03 13:40:04 +0200 CEST  |       ->   6,3          [anomalous CPU spike: CPU 22117.000000 > average 22115.000000] 
2019-06-03 13:40:04 +0200 CEST  |    6 -->   7,1      A sideline to test some raw concept mapping 
2019-06-03 13:40:04 +0200 CEST  |       ->   7,2        [btw: Commence testing] 
2019-06-03 13:40:04 +0200 CEST  |    7 -->   8,1      End of sideline concept test 
2019-06-03 13:40:04 +0200 CEST  |       ->   8,2        [btw: Commence testing] 
2019-06-03 13:40:04 +0200 CEST  |    6 go>   9,1      Starting Kubernetes deployment 
2019-06-03 13:40:04 +0200 CEST  |       ->   9,2        [btw: Commence testing] 
2019-06-03 13:40:07 +0200 CEST  |       ->   9,3          [remarked: : Starting kubernetes pod] 
2019-06-03 13:40:10 +0200 CEST  |       ->   9,3          [remarked: : File drop in pipeline] 
2019-06-03 13:40:13 +0200 CEST  |       ->   9,3          [remarked: : Querying data model] 
2019-06-03 13:40:16 +0200 CEST  |       ->   9,3          [remarked: : Submit transformation result] 
2019-06-03 13:40:19 +0200 CEST  |    6 go>  10,1      The end! 
2019-06-03 13:40:19 +0200 CEST  |   10 -->  11,1      Show the signposts 
\end{alltt}
\caption{\small An expanded view of a compressed local checkpoint log
  for a process. The first column is system clock timestamp which
  provides approximate time of day context for relating events to
  human scales.  The next fields use interior monotonic counters that
  increment on SignPost events, even through concurrent coroutines.
  Each event points back to the preceding event, to give actual causal
  history. Explanations of the event semantics are shown on the right.\label{map1}}
\end{center}
\end{figure*}

\subsection{Dropping hints}

To record useful events, from within the meaningful context of a
process, processes need an API that constrains authors to produce information
that can be consumed later.  For example, in the Koalja history
package, based on the principles in this paper, significant events can
be marked with signposts\cite{koaljahistory}.
\begin{alltt}
\small
H.SignPost(&ctx,"Milestone 1...")
H.SignPost(&ctx,"Milestone 2...")
...
H.SignPost(&ctx,"Commence testing")
\end{alltt}
And these signposts can be detailed, using the four spacetime
semantic relations from \cite{spacetime3,cognitive}:
\begin{alltt}
\small
H.SignPost(&ctx,"code signpost X").
    Intent("open file X").
    ReliesOn(H.NR("/etc/passed","file")).
    FailedBecause("xxx").
    PartOf(H.NR("main","coroutine"))

\end{alltt}
We shall explain this point further in a sequel\cite{william2}.

\section{Model extraction}\label{model}

If we pursue a concrete strategy of separating timescales and
extracting invariants from the chaff of noisy variation, we can
expect to infer causal and conceptual relationships over long
aggregate times by learning.  Learning is a process that happens across several
timescales, as noted in \cite{spacetime3}. In modern Machine Learning
parlance, we would say that acquiring and stabilizing training data is
a long term process, while recognition and classification is a short
term process. Monitoring tries to achieve both processes
as an unsupervised in band single-scale process, so it has to deal with
the instabilities in band too.

The spacetime model in \cite{spacetime3,cognitive} allows us to define
a partial ordering of semantics, represented as agents in a virtual
knowledge space.  The future and past cones are generated by the first
two spacetime semantic relations: for generalization or scope and
causal order.  Ordered relationships are the most important ones
because they tell the stories we seek. Data may arrive in incidental order,
for a variety of reason that involve causal mixing. We need to extract
the intended order of system cooperation from the incidental or unintended
order of side channels that muddle behaviour.

\subsection{Invariant sequences form explanations}

In order to generate a map of invariants, we need to not only identify them,
but consider what they can promise about one another, so that we may position them
in relative terms.
In \cite{spacetime3}, it was shown that we can plausibly define four kinds of
semantic relationship based on elementary spacetime considerations. These ought to
apply between pairs of agents in any distributed system.
\begin{itemize}
\item i) Contains: localization in spacetime (scope of containment or ordering by scale)
\item ii) Follows: order, causation (Markov processes or order by influence)
\item iii) Expresses: local distinction (scalar attribute at any scale)
\item iv) Near to: measure, distance (assessment of distance at any scale)
\end{itemize}
Notice that order is distinct from distance, i.e. direction and proximity
are different concepts. These concepts are not clearly distinguished in a vector space.

\begin{figure}[ht]
\begin{center}
\includegraphics[width=6cm]{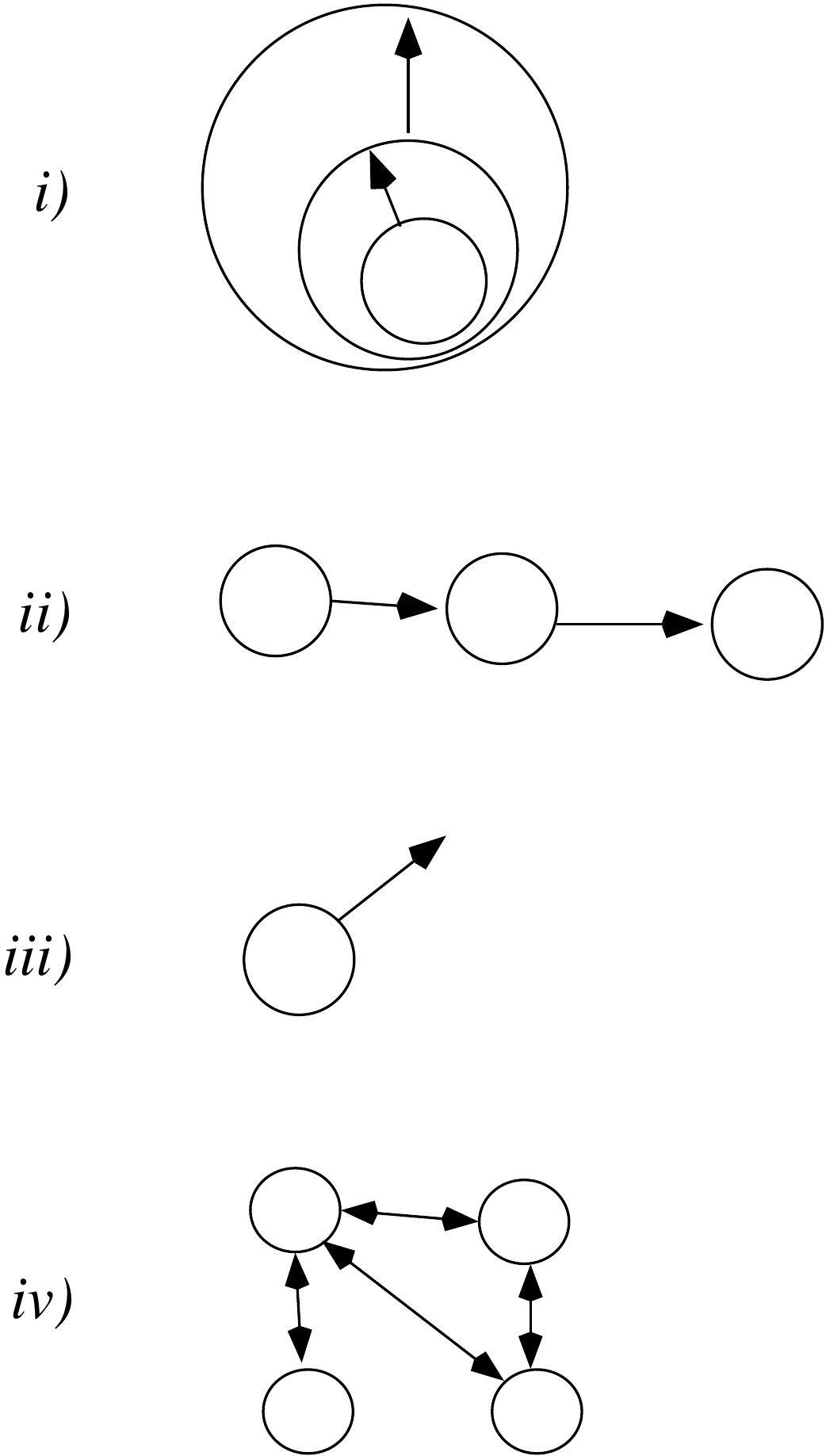}
\caption{\small The four kinds of promise that spacetime can express:
  i) containment, ii) succession, iii) local attributes, and iv)
  proximity. Although we can distinguish different sub-types of these
  four, it's hypothesize that the four are necessary and sufficient
  for describing observable phenomena.\label{fourST}}
\end{center}
\end{figure}

Agents may promises exemplifiers, symbolic and metric discriminators, or regional classifiers.

The relationship between a discriminator iii) and a classifier iv) is
subtle, and is easy to promise inconsistently. The essence of
expressing an attribute is to label the type, which can be
combined with something else to form a union of different promises
(a kind of semantic chemistry).  A promise of containment, on the
other hand, is a promise of belonging to a common class.  These two
promises therefore represent assembly versus classification of agents.

The extent to which we have the ability to localize a causal influence is the essence of `root
cause' analysis: the ability to contain the process within a virtual
boundary which itself can make promises on a new level.
This is part of the motivation for virtualization and containerization.

\begin{figure*}[ht]
\begin{center}
\begin{alltt}
\footnotesize
<begin NON-LOCAL CAUSE>
(program start) --b(precedes)--> "MainLoop start"
  (MainLoop start) --b(precedes)--> "Beginning of test code"
     (Beginning of test code) --b(precedes)--> "code signpost X"
        (code signpost X) --b(precedes)--> "Run ps command"
        (code signpost X) --b(precedes)--> "TEST1---------"
        (code signpost X) --b(precedes)--> "Commence testing"
     (Commence testing) --b(precedes)--> "The end!"
        (Commence testing) --b(precedes)--> "[remarked: : Possibly anomalous CPU spike for this CPU]"
        (Commence testing) --b(precedes)--> "A sideline to test some raw concept mapping"
        (Commence testing) --b(precedes)--> "Starting Kubernetes deployment"
        (Commence testing) --b(precedes)--> "The end!"
        (Commence testing) --b(precedes)--> "[remarked: : Possibly anomalous CPU spike for this CPU]"
        (Commence testing) --b(precedes)--> "A sideline to test some raw concept mapping"
        (Commence testing) --b(precedes)--> "Starting Kubernetes deployment"
     (The end!) --b(precedes)--> "Show the signposts"
        (A sideline to test some raw concept mapping) --b(precedes)--> "End of sideline concept test"
     (The end!) --b(precedes)--> "Show the signposts"
        (A sideline to test some raw concept mapping) --b(precedes)--> "End of sideline concept test"
     (code signpost X) --b(may determine)--> "[dns lookup: 123.456.789.123]"
        (TEST1---------) --b(may determine)--> "[file: file://URI]"
        (TEST1---------) --b(may determine)--> "[file: file://URI]"
<end NON-LOCAL CAUSE>
\end{alltt}
\caption{\small An excerpt of a map of invariants, generated by a search.
Invariants are accumulated from distributed and concurrent processes
and their relationships are classified by the four spacetime types. The pathways
through these relationships tell different kinds of stories. The excerpt
shown involves expansive reasoning: combining generalization and causality.\label{map2}}
\end{center}
\end{figure*}

\subsection{Promising semantic maps}

The four spacetime semantic relationships, described in
\cite{spacetime3,cognitive} may be assigned between pairs of concepts,
originating by signal (+) or by inference (-), entirely at the behest
of an observer, and  according to the following `selection rules':

\begin{enumerate}
\item {\bf Distinguishability}: Descriptive properties that
  distinguish, describe, and embellish the name of a concept are {\bf
    EXPRESS} promise types. These are scalar promises used to explain
  attributes that may form compositions of attributes for aggregate
  `hub concepts'.

  For example: a banana may express the colour yellow, ripeness, and
  sweetness. It does not express fruit or Del Monte.

\item {\bf Generalization}: membership in classes and informal
  categories use the {\bf CONTAINS} promise type. These express
  subordination to one or more umbrella concepts, and superordination
  to instances and exemplars of the named concept. Generalization is
  strictly transitive.

  Generalization is not as in taxonomy: a concept may have any number
  of generalizations, i.e. there is no unique typology to concepts.
  The utility of recognition lies in the overlapping nature of
  classes\cite{tiny}.

  For example, a banana is generalized by fruit and desserts, and has
  instances such as Del Monte.  It does not express these as
  attributes.

\item {\bf Dependency}: promises of dependency---prerequisite or
  follow-up concepts are {\bf FOLLOWS} type promises. They may link
  concepts of any type into some meaningful order, by any
  interpretation of the observer.

  For example, the beginning precedes the end. ``One'' precedes
  ``two'' which precedes ``three'', etc.  Dependency is usually
  transitive, but may contain loops (in feedback cycles).

\item {\bf Similarity}: the degree of similarity between two concepts
  is an assessment that may be promised by any observer, to represent
  a degree of similarity or closeness. This is represented by promises
  of type {\bf NEAR}. This is an ad hoc assessment and should not be
  taken too seriously.

  In the case where several agents form an agreement about the metric
  distance relationships between concepts, these assessments may form
  the basis for a shared local coordinate system.
\end{enumerate}

\begin{figure}[ht]
\begin{center}
\includegraphics[width=6cm]{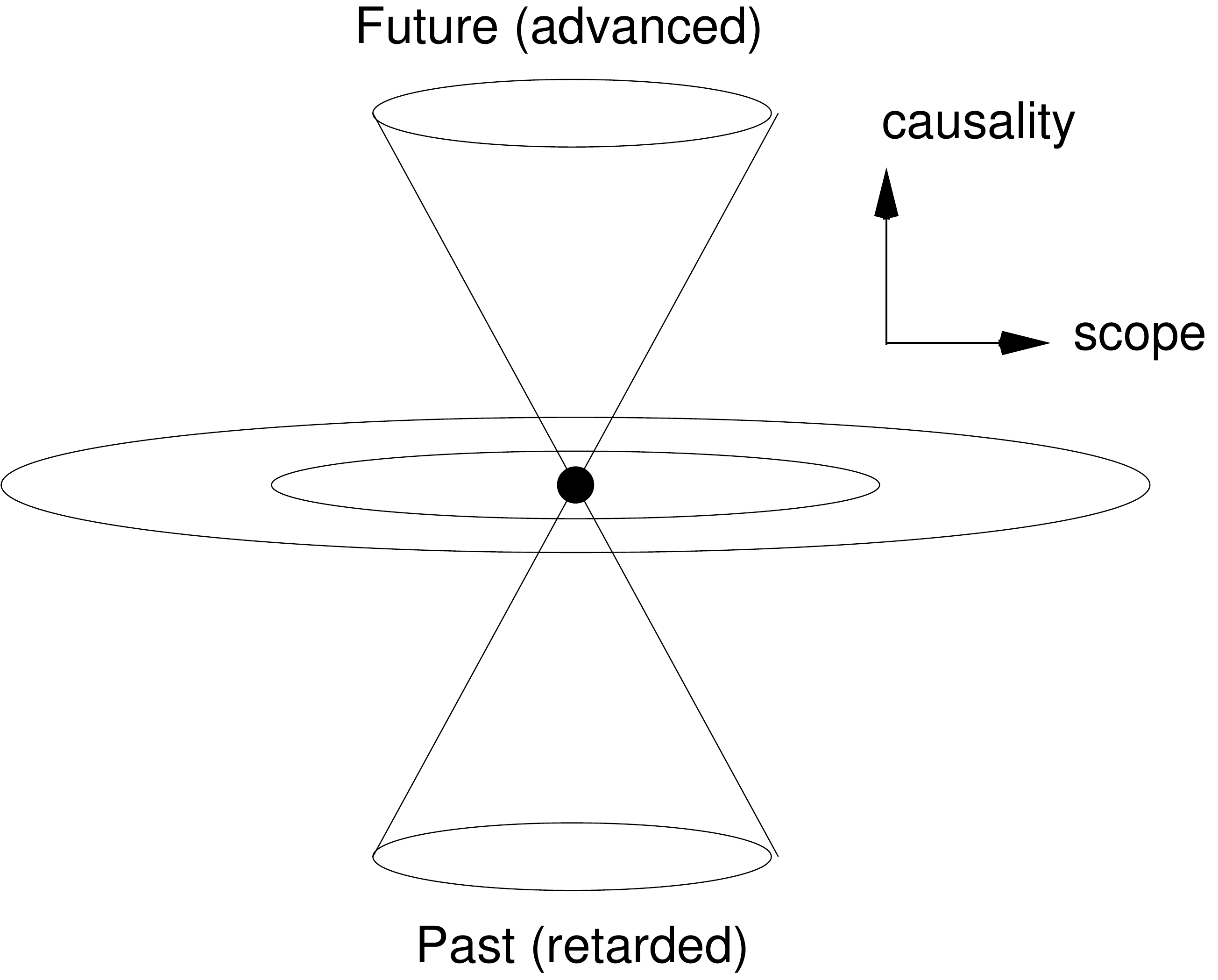}
\caption{\small The propagation cones indicate the past and future as `timelike'
trajectories generated by the causal relationship, and semantic scale or scope
of meaning accumulated in `spacelike' directions around the average time axis
in rings of increasing generalization.\label{cones}}
\end{center}
\end{figure}

The semantics of these relationships are not automatically orthogonal to one another,
so we have to maintain the incompatibility of the types by assignment. The local
promises are mutually incompatible, which is to say the no two agents may
promise more than one of the three types {\bf CONTAINS, FOLLOWS, EXPRESSES} (or
their inverses\footnote{We must be cautious and pay attention to the Promise Theory
principle that just because one agent promises to contain another or be followed by another,
it does not imply that the agent concerned agrees with this, and may not promise it.
For example, firewalls may create a one-way glass effect that prevents
the inverse from being implemented)}.
\begin{itemize}
\item {\bf EXPRESSES} is incompatible with {\bf CONTAINS}.
\item {\bf CONTAINS} is incompatible with {\bf FOLLOWS}.
\item {\bf FOLLOWS} is incompatible with {\bf EXPRESSES}.
\end{itemize}
The semantics are easily illustrated with an example. The concepts blue and yellow
are expressed by objects that combine them as part of their identity:
e.g. a blue and yellow pattern, like the Swedish national flag, and green paint
may be composed of blue and yellow paint, but neither the Swedish flag nor green
are generalizations of blue and yellow. The concept of colour, on the other hand
does not express blue or yellow, but generalizes them as members.

The promise of proximity is slightly different:
\begin{itemize}
\item {\bf NEAR} is potentially compatible with any of the above, since it is an informal
assessment of non-locality.

The assessment of proximity between agents may seem to imply something about the
orthogonal semantics above, but this is ambiguous (see figure \ref{near}).
For example, because proximity is a type of relationship, not a standardized metric constraint,
relations may vary in their interpretation:

\beq
&A \text{~\rm\bf EXPRESSES~} \text{`close to B'}&\\
&A \text{~\rm\bf EXPRESSES~} \text{`close to C'}&\\
&\text{`close to B'} \text{~\rm\bf FOLLOWS~} \text{`close to C'}~~&
\eeq
Together these might suggest that $A,B,C$ all lie in a certain region and that there
must therefore be a category (dotted line in figure \ref{near}) that generalizes all of them.
That kind of inference is dangerous, because it is based on coarse inference, 

\begin{figure}[ht]
\begin{center}
\includegraphics[width=4cm]{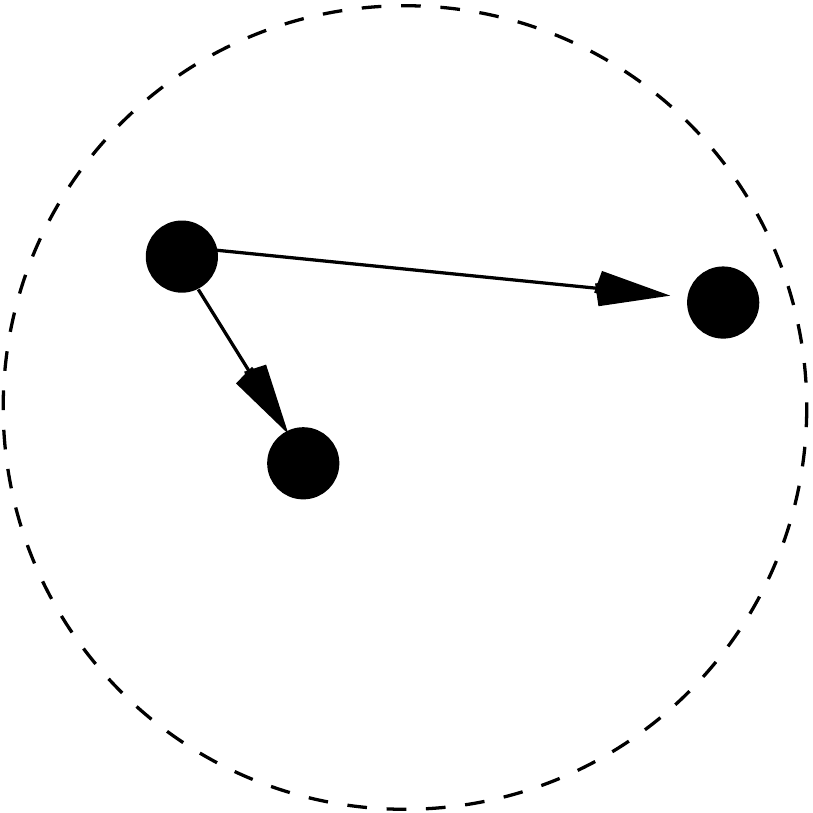}
\caption{\small The assessment of proximity between agents may seem to imply something about the
orthogonal semantics above, but this is ambiguous.\label{near}}
\end{center}
\end{figure}

\end{itemize}
These selection rules can be applied in order to join similar objects into hubs.
Each observed instance maps to a hub that can be broken down into atomic concepts
by expression. Containment promises are generally learned on a much longer timescale (e.g.
added by human expertise), and causal dependency promises are added by processes that
generate them or observe them.

\subsection{Storytelling from spacetime semantics}

Once constructed, the graph may be parsed to generate stories, or
automated reasoning.  A reasoning process may be viewed as an
expansive search along alternating (+) axes (causal outcomes that are
related by generalization or exemplification by specific instance),
and tempered by elimination by relevance criteria (-).

\begin{figure}[ht]
\begin{center}
\includegraphics[width=6cm]{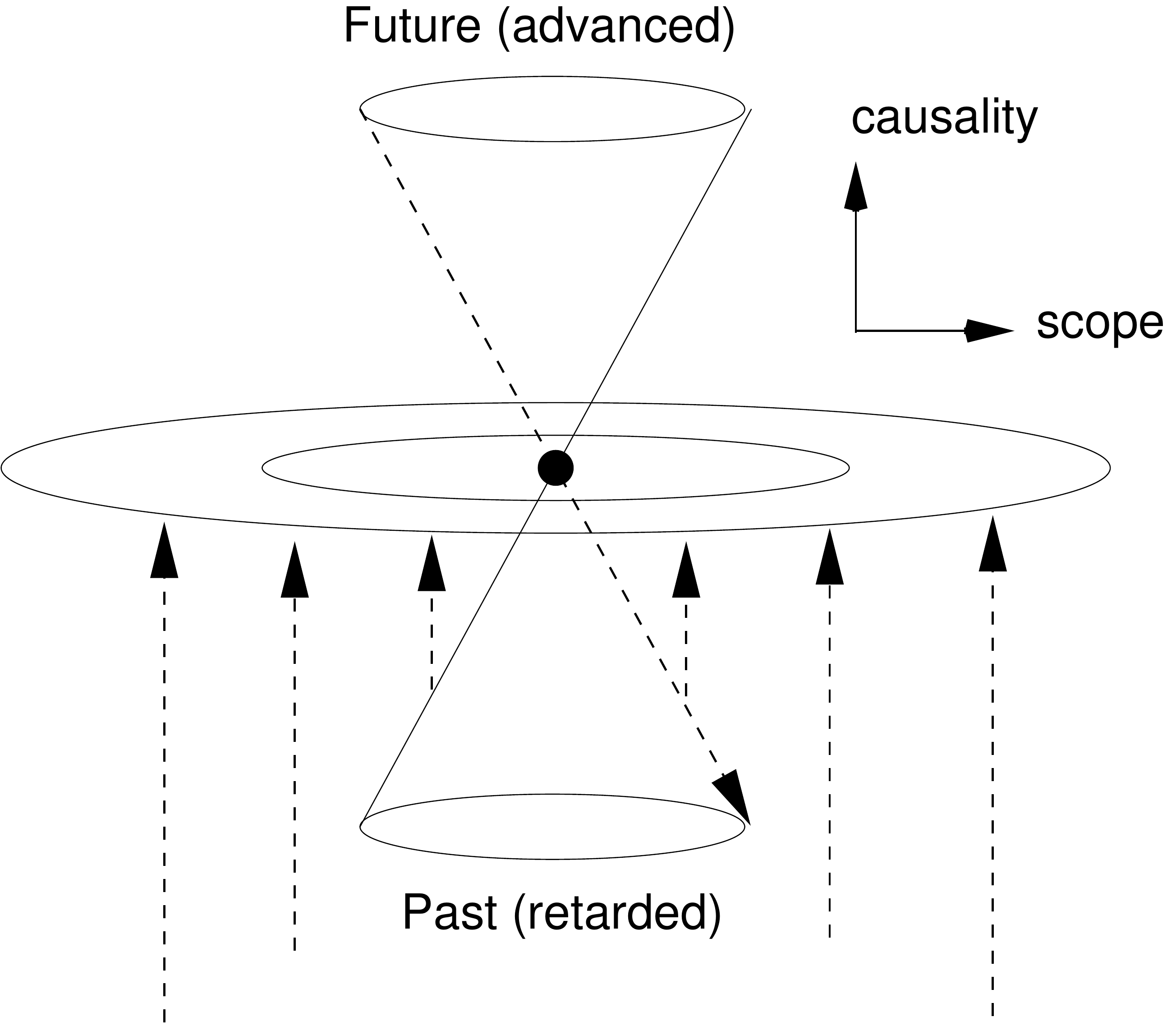}
\caption{\small The scope of knowledge about spacelike information is accumulated as memory from past events propagated into
a model of the present.\label{cones2}}
\end{center}
\end{figure}

\begin{itemize}
\item Starting from a topic of interest, we follow promises of type {\bf FOLLOWS} independently
in the forward and backward directions, to explore the causal cone (figure \ref{cones2}).

\item Arriving at each new concept, we follow promises to generalize and specialize the concept
to find all links arising from the collective generalized concept, and follow these
along different story paths. In other words, we multiply the number of stories
by conceptual associations that imply examples of the same idea---expanding the scope
of meaning without going off the rails.
\end{itemize}
Such promised relationships cannot be easily found by in band machine learning techniques: this
is an orthogonal and complementary method, but learning may form the basis for collapsing
experience into an arrangement of similar concepts on longer timescales (see figure \ref{cones2}).

The algorithmic rules for parsing stories from the concept graph come in
several forms.  A conceptual reasoning search might be bounded at the start
(retarded), at the end (advanced) or at both ends (causal).
The first two are a form of brainstorming that ends with a single concept:
`tell me all about X'. The latter case asks for a specific explanation
of `Y given X'. There is no unique path for any search, in general.
The paths most frequently trodden, i.e. have the most frequently observed transitions,
or most frequently searched for concepts, become `classical paths'
and may be favoured, someone analogous to a PageRank search\cite{pagerank,graphpaper}.

Loops in causal relationships may be significant, so we should detect them.
Some loops may be errors of identification, others may be cyclic reasoning (e.g.
self-consistent ideas, like eigenvalue problems).

\section{Models, sharding, idempotence, and forgetting}

From sampling of data at the edge of a network, to actionable insight, there is a
chain of reasoning to monitoring that starts with observability and ends with the deletion
of irrelevant and antiquated data:
\begin{enumerate}
\item Data collection,
\item Stability or convergence to fixed points,
\item Model extraction,
\item Classification into buckets,
\item Controlled forgetting.
\end{enumerate}
Each of these steps plays an important role.  Data collection provides
the basic observability to trace systems at different scales and tell
stories about them that bring valued insights. The convergence of
phenomena to fixed points is an incredibly important principle in
dynamics, but one that receives too little attention\footnote{This is a
  principle that I have reiterated many times since
  CFEngine\cite{burgessC1} to stabilize and guide us towards invariant
  meaning.}. When systems fly all over the place, they are not telling
us anything significant. It's only when they converge onto repeated
patterns or stable attractors that we can build on them as part of a
reliable. Models need to expose those differences. Today, there is a
fascination with using machine learning to try to expose such fixed
points, but the technique is only possible if there is sufficient
stability. When monitoring reaches a level of maturity in IT, we will
place as much value in attending to semantics as we do in recording
noise today.
Data that have the same semantics need not be recorded twice. Once one
has identified the invariants of a system, these can be made
idempotent. Model identification classifies inputs into discrete
alphabets. Repeated symbols can be compressed, and if they are repeated
no harm need be done if they have fixed point semantics.

For example, it is not a problem if we accidentally collect the same
data twice as long as they map to the same place. Storing the same
data twice is idempotent unless we are counting frequencies. Frequency
counting can be made idempotent by labelling intervals. The general
principle is that we should engineer data sources to permit
convergence. Promise Theory reveals the mutual responsibility for
information transfer between sender and receiver.
\begin{principle}[Convergent data]
The safest way to avoid data inconsistency is to design messages
in such a way that repeated messages always map to the same location
and update them without breaking a promise.
\end{principle}

Fixed points lead to stable models, which lead to
efficient indexing of knowledge. This helps in scaling storage (e.g.
in sharding), and it helps in fault tolerance. When data are recorded
around proper index points, it doesn't matter if data get delivered
multiple times (idempotence) or even out of order: everything will
find its proper place in the end.  Model extraction tells us how to
compress the data into an alphabet or catalogue of meaningful and
significant ideas, and therefore separate into buckets or shards.

Finally, perhaps the most important issue of all is how to forget what
is no longer of value.  Keeping data and even models around forever is
a senseless squandering of resources and an irresponsible and
unsustainable use of technology. One wonders how many of the photos
now being eagerly accumulated in the cloud will be preserved in ten
years' time. The same is true of monitoring data that were collected
last week. If we don't understand the timescales, context, and
relevance of data, then we have no business collecting it, because it
cannot tell us anything of value.  A policy for forgetting can usually
be built into a definition of context from the start, e.g. through finite windows and
sliding sets, running averages, and so on \cite{burgessDSOM2002}.

\begin{example}[Model based collection]
In CFEngine, weekly data were mapped idempotently to a finite
number of buckets marked by 5 minute intervals throughout a week, based
on a prior measurement survey using autocorrelations.
After a weekly period, the buckets would wrap around, like a clockface
and data would update the corresponding image in the map. This
approach was able to promise a limited stability of expectations, as
well as automated forgetting (constant weight gradient of temporal history),
and thus effective garbage collection
\end{example}

What's remarkable is how few of these issues actually get any
attention in the literature. One hears arguments like `its cheap to
keep all data forever'---which smacks of the sudden realization of
global warming or the plastic crisis. Every time we advocate
increasing something, we need to think about the balancing garbage
collection process.

\section{Summary and conclusions}

The collection of accurate data is not in question. Today, there is
increasing interest capturing `digital twin' representations of agents
in the real world, with every detail available for possible
inspection. No one could resist the idea of such knowledge, unless it
invades their privacy. The question in technology monitoring is rather
whether every detail should be centralized and whether data can be
compressed without loss.

In this summary of what can be observed about distributed systems, we
see that tracing events back to `root cause' is an ill-defined
problem, but tracing back a significant likely cause is indeed
possible with careful labelling (especially of time). This kind of labelling is not
commonly provided in current tooling.
Assuming access to data,
we might hypothesize that a complete monitoring system would promise to:
\begin{itemize}
\item Separate timescales.
\item Identify the alphabet of system invariants
\item Capture local histories of instances, in the context in which they happen.
\item Identify significant events at different scales and measure their invariance.
\item Tools for reconstructing and backtracing of histories from local data.
\item Tools for generating semantic past-future cones for causal reasoning.
\end{itemize}
Today, many IT monitoring systems transmit raw data in large
quantities to a central point for analysis, without attempting to
alphabetize the data before transmission. In effect, by ignoring the
existence of a model (summarized by an alphabet of non-overlapping
signals) one is repeatedly sending the same model over the network
again and again, wastefully, and to no gain. If we can classify observations at their source,
and condense them into an alphabet of signals, a vast data compression
can be accomplished for both faster recognition and potentially
greater semantic content. That will be the subject for a sequel \cite{william2}.

It should be clear that nothing about the ability to trace systems
enables full reversibility of state, which should be considered
difficult to impossible, depending on scale\cite{jan60}---so ultimately
monitoring may be of little value. More value could be captured by
building intrinsic stability into systems in the first place.

The elephant in the monitoring system is an essential attitude in the
industry concerning the purpose of monitoring. Systems are only
sustainable, knowable, and predictable when they seek
stability---not when they labour under the burden of intrusive
inspection. For a lot of practitioners there is a conflict of interest
here. If we seek to measure all that is random or unstable, by
oversampling and consuming resources wastefully, it will be neither
stable nor sustainable.  Consensus protocols, for instance, promise
semantic stability, and are popular (if somewhat over-used) in
software engineering\footnote{The question of whether to invest in
  promising an expensive and late consensus over a coarse grain of
  space, or whether to expose its divergences as a feature remains a
  policy choice---one that currently aligns with opposite poles of Dev
  and Ops.}. They draw attention to a preoccupation with semantics in
software engineering, i.e. a desire for stable qualitative outcomes,
at the expense of quantitative delay.  Software engineers seem not to
trust concepts like intrinsic dynamical stability i.e.  systems that
promise to converge to predictable quantitative outcomes.  Monitoring
tends to treat all software as adversarial, and we put more faith in
ill-designed monitoring than in an initial software design.  This is a
paradox that will inevitably lead to big surprises and catastrophic
events.

By focusing on essentials, there are many issues I've not had time to
mention in this paper.  I hope to return to some of these in future
work.

{\bf Acknowledgements:} I am especially grateful to William Louth for discussions.
I'd also like to thank Nicolas Charles, Simon Lucy, Colm MacC\'arthaigh,
and Adrian Cockcroft for helpful comments and references.

\bibliographystyle{unsrt}
\bibliography{spacetime}

\end{document}